\documentclass[final, twocolumn]{svjour3}
\usepackage[utf8]{inputenc}
\usepackage[english]{babel}
\usepackage{cuted}
\usepackage{amsmath}
\usepackage{amsfonts}
\usepackage{amssymb}
\usepackage{graphicx}
\usepackage{listings}
\usepackage{caption}
\usepackage[square,sort,comma,numbers]{natbib}

\usepackage{color}
 
\definecolor{codegreen}{rgb}{0,0.6,0}
\definecolor{codegray}{rgb}{0.5,0.5,0.5}
\definecolor{codepurple}{rgb}{0.58,0,0.82}
\definecolor{backcolour}{rgb}{0.95,0.95,0.92}
 
\lstdefinestyle{mystyle}{
    backgroundcolor=\color{backcolour},   
    commentstyle=\color{codegreen},
    keywordstyle=\color{magenta},
    numberstyle=\tiny\color{codegray},
    stringstyle=\color{codepurple},
    basicstyle=\footnotesize\ttfamily,
    breakatwhitespace=false,         
    breaklines=true,                 
    captionpos=b,                    
    keepspaces=true,                 
    numbersep=5pt,                  
    showspaces=false,                
    showstringspaces=false,
    showtabs=false,                  
    tabsize=2
}
 
\begin{document}
	
\title{Solcore: A multi-scale, Python-based library for modelling solar cells and semiconductor materials}
\titlerunning{Solcore}      

\author{D. Alonso-\'Alvarez   \and T. Wilson \and P. Pearce \and M. F\"uhrer \and D. Farrell \and N. Ekins-Daukes}

\authorrunning{D. Alonso-\'Alvarez \textit{et al.}} 

\institute{D. Alonso-\'Alvarez \at
              Department of Physics, Imperial College London, London, United Kingdom \\
                            Tel.: +44 (0) 20 759 47563\\
              \email{d.alonso-alvarez@imperial.ac.uk}            \\
              \emph{ORCID:} 0000-0002-0060-9495  \\
           \and
           T. Wilson \at
              Department of Physics, Imperial College London, London, United Kingdom \\
           \and
           P. Pearce \at
              Department of Physics, Imperial College London, London, United Kingdom \\
           \and
           M. F\"uhrer \at
              Department of Physics, Imperial College London, London, United Kingdom \\
           \and
           D. Farrell \at
              Department of Physics, Imperial College London, London, United Kingdom \\
           \and
           N. Ekins-Daukes \at
              Department of Physics, Imperial College London, London, United Kingdom \\
 			\emph{Present address:} School of Photovoltaic and Renewable Energy Engineering, University of New South Wales, Sydney, Australia \\
              }

\journalname{Journal of Computational Electronics}
\date{Received: 27th November 2017 / Accepted: date}

\maketitle

\begin{abstract}

Computational models can provide significant insight into the operation mechanisms and deficiencies of photovoltaic solar cells. Solcore is a modular set of computational tools, written in Python 3, for the design and simulation of photovoltaic solar cells. Calculations can be performed on ideal, thermodynamic limiting behaviour, through to fitting experimentally accessible parameters such as dark and light IV curves and luminescence. Uniquely, it combines a complete semiconductor solver capable of modelling the optical and electrical properties of a wide range of solar cells, from quantum well devices to multi-junction solar cells. The model is a multi-scale simulation accounting for nanoscale phenomena such as the quantum confinement effects of semiconductor nanostructures, to micron level propagation of light through to the overall performance of solar arrays, including the modelling of the spectral irradiance based on atmospheric conditions. In this article we summarize the capabilities in addition to providing the physical insight and mathematical formulation behind the software with the purpose of serving as both a research and teaching tool. 

\end{abstract}


\section{Introduction}

Computer aided design and device models are valuable tools when developing and evaluating photovoltaic solar cells. Laboratory scale tests can be usefully compared against detailed models that account for all relevant processes or with ideal, thermodynamically limited behaviour. Over the years, and with different degrees of sophistication, many pieces of software have been developed and published to tackle different aspects of solar energy research. For example, to calculate the solar spectrum as a function of the atmospheric conditions a traditional solution is to use SMARTS~\cite{Gueymard:1995tp}; the light absorption profile in the solar cell or even at module level could be addressed by OPTOS~\cite{Eisenlohr:2015ega} or OPAL2~\cite{BakerFinch:2010ho}; while to solve the transport equations of a solar cell one could use PC1D~\cite{Basore:wP5Wjg}, SCAPS~\cite{Burgelman:2000cd} or Quokka~\cite{Fell:2013dq}. Several free and commercial programs, not specifically designed for solar energy research, have also been used historically, including AFORS-HET~\cite{VARACHE201514}, Nextnano~\citep{Birner:2007gm}, ATLAS~\cite{AtlasDeviceSimula:vx} or SENTAURUS~\cite{Sentaurus}, with the first two focused on the device and semiconductor properties and the latter two also solving the optics of the solar cells, among other properties. An extensive list of software for solar energy research - both online calculators and downloadable programs - has been compiled by PV Lighthouse~\cite{PVLighthouse}. In general, programs like ATLAS and SENTAURUS provide a general-purpose, easy to use interface - often solving multi-physics problems, such as electrical transport coupled with thermal transport - to the detriment of performance. On the contrary, specific programs like AFORS-HET or PC1D are extremely fast and efficient, but limited in the problems they can solve, in this case 1D heterostructures and solar cells. 

Apart from a few exceptions, such as PVlib~\cite{Andrews:2014fu}, all these solvers are high-level, self-contained applications. While users can provide their own inputs and, in some cases, access the source code of the programs and customize some aspects of them, they are not designed with that purpose in mind. 

Solcore is a multiscale, modular simulation framework for solar energy research, written mostly in Python. Solcore evolved from SOL, a Fortran-based, quantum well solar cell solver developed by Nelson and Connelly~\cite{Connolly2012}, with the explicit purpose of simplifying its integration in other programs, its expansion with custom routines and algorithms, and being didactic and informative. It is a teaching and learning tool as much as a rigorous research tool. Solcore is also extremely flexible. It integrates several algorithms for the multiscale modelling of semiconductors and solar cells, allowing a user without access to other methods, but with some experience with Python, to solve many different problems out of the box. On the other hand, most of its functionality can be interfaced with external tools, optimized to solve a specific problem, which are more advanced and accurate or that use an approach not considered in Solcore. The most recent version has been released under the GNU Lesser General Public License (GNU-LGPL) and can be found on GitHub~\cite{SolcoreCode}.  

\begin{figure*}
	\centering
  	\includegraphics[width=\textwidth]{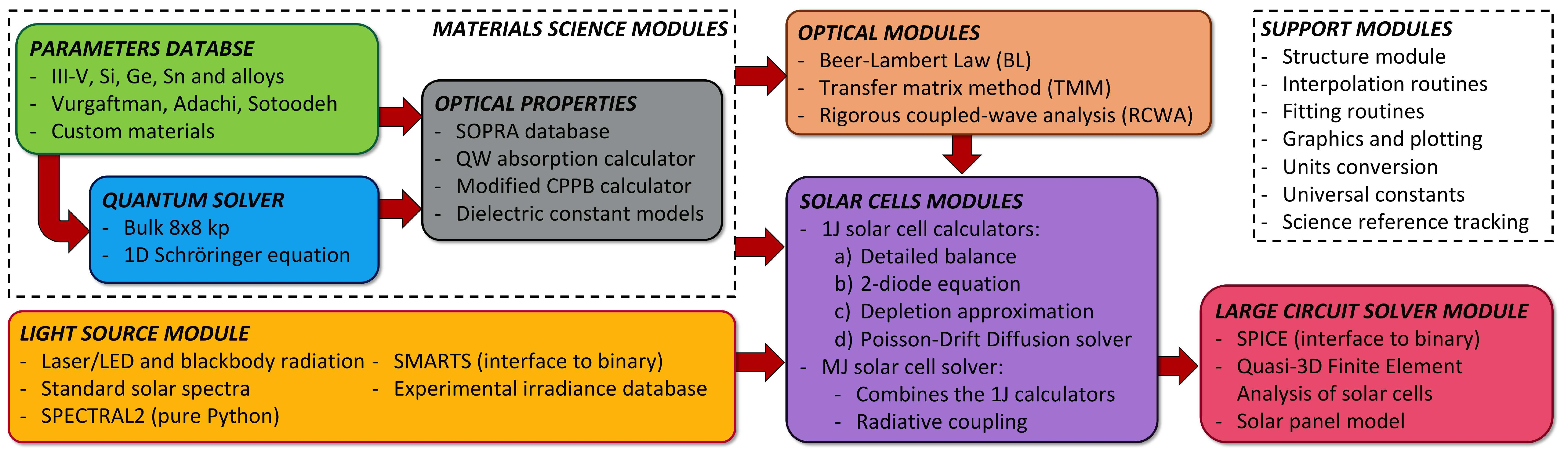}
  	\caption{General structure and workflow of Solcore.}
  	\label{fig:solcore_overview}
\end{figure*}

Solcore's capabilities can be grouped into four categories: materials science (Section \ref{sec:materials science}), light sources (Section \ref{sec:light sources}), solar cells (Sections \ref{sec:Optical solvers} to \ref{sec:MJ}) and large-scale calculators (Section \ref{sec:large scale}), each of them tackling a different area and scale relevant for research in solar energy. Figure \ref{fig:solcore_overview} shows how these parts relate to each other and summarizes some of their content. 

\section{Materials science} \label{sec:materials science}

The materials science modules in Solcore deal with the retrieval and calculation of material properties as well as those of quantum nanostructures, in particular quantum wells. They form the building blocks necessary to create the structures and calculate the performance of full solar cell devices. While focused on its application for solar cells, this part of Solcore is widely applicable to any research area related to semiconductor materials, as a way of managing the material properties, customising them and using them in other calculations.

It should be noted that, in reality, electronic and optical properties are not independent but related to each other via the material band structure. As the case in most programs, Solcore uses a non-consistent approach, with independent electronic and optical parameters obtained from different sources. The reader should consider full band structure methods, like pseudopotential or tight-binding, for a consistent set of electronic and optical properties.  

\subsection{Parameters database}

The parameters database contains the basic properties of many semiconductor materials, including silicon, germanium and many III-V semiconductor binary and ternary alloys. Among other parameters, it includes the energy bandgap, the electron and hole effective masses, the lattice constants and the elastic constants. 

The main sources of data are the article by I. Vurgaftman on \textit{Band parameters for III-V semiconductors} \cite{Vurgaftman:2001bu} and the \textit{Handbook Series on Semiconductor Parameters} by Levinshtein et al. ~\cite{Levinshtein:2012fv}. The carrier mobility calculator is based on the empirical low-field mobility model by Sotoodeh et al.~\cite{Sotoodeh:2000jq} and it is available only for some materials where the inputs for the model are available. 

There are two methods for retrieving parameters from the database. The first one consists simply of getting the data using the \texttt{get\_parameter} function with the required inputs. For example:

\begin{lstlisting}[language=Python]
get_parameter("GaAsP", "band_gap", P=0.45, T=300)
\end{lstlisting}

\noindent will return the bandgap of GaAsP for a phosphorus concentration of 45\% at a temperature of 300 K, equal to 1.988 eV. This method only uses the existing data. Another method is to create a material object which will contain all the properties existing in the database for that material, as well as those included as input, which will override the value of the database parameter, if it exists. The following example creates a GaAs object and an AlGaAs object, using a custom electron effective mass in the latter: 

\begin{lstlisting}[language=Python]
GaAs = material("GaAs")(T=300, Na=1e24)
AlGaAs = material("AlGaAs")(T=300, Al=0.3, Nd=1e23, eff_mass_electron=0.1)
\end{lstlisting}

Now, any parameter - including the custom ones - are attributes that can be easily accessed and used anywhere in the program. For example \texttt{GaAs.band\_gap} is the GaAs bandgap and \texttt{AlGaAs.lattice\_constant} is the AlGaAs lattice constant, both at the composition and temperature chosen when creating the objects. 

\begin{figure}
	\centering
  	\includegraphics[width=\columnwidth]{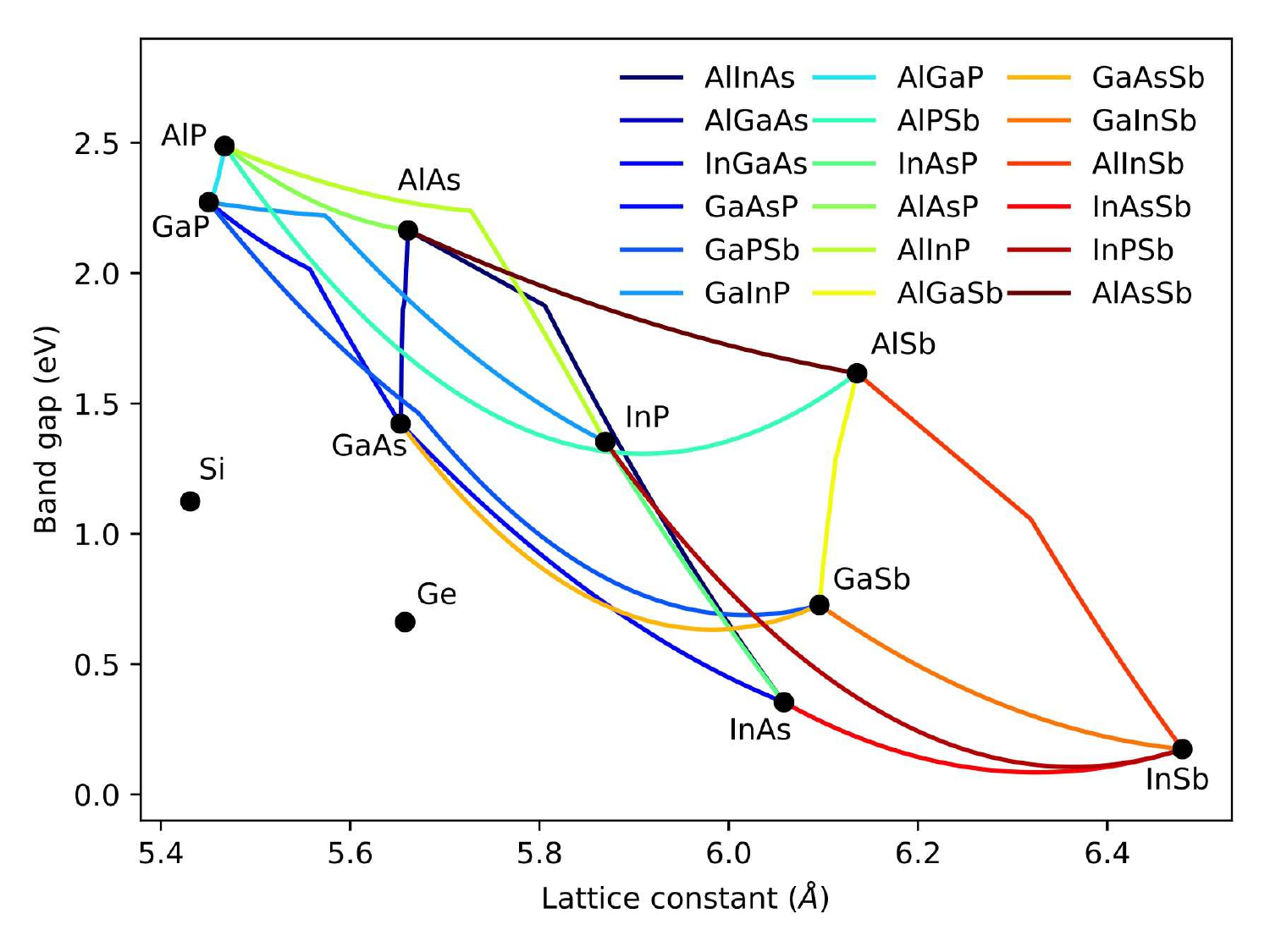}
  	\caption{Materials with most parameters included in Solcore's database (excluding the optical properties).}
  	\label{fig:materials}
\end{figure}

Figure \ref{fig:materials} shows the well-known bandgap vs. lattice constant map of all semiconductor materials and alloys (only ternary compounds) currently implemented into Solcore. However, any other material can be used in all of the Solcore functions, as long as the necessary input parameters are provided. This can be done by overriding all the properties of an existing material during the creation as above, or adding it as an external material in the configuration files. 

\subsection{Optical Properties Database}

In order to calculate and model the optical response of potential solar cell architectures and material systems, access to a library of accurate optical constant data is essential. Therefore, Solcore incorporates a resource of freely available optical constant data measured by Sopra S. A. and provided by Software Spectra Inc. \cite{sspectra_2008}. The refractive index $n$ and extinction coefficient $k$ are provided for over 200 materials, including many III-V, II-VI and group IV compounds in addition to a range of common metals, glasses and dielectrics. 

Any material within the Sopra S. A. optical constant database can be used with the ``material'' function described above, but they will have only the optical parameters $n$ and $k$. In the case of materials that are in both databases, the keyword ``sopra'' will need to be set to ``True'' when creating the material. Once a material is loaded its $n$, $k$ and absorption coefficient data is returned by calling the appropriate method, for example \texttt{SiO2.n(wavelength)} and \texttt{SiO2.k(wavelength)}. For certain materials in the database, the optical constants are provided for a range of alloy compositions. In these cases, any desired composition within the range can be specified and the interpolated $n$ and $k$ data is returned. Several examples of materials created from the Sopra database are shown in the Listing \ref{lst:sopra}. 

\begin{lstlisting}[language=Python, caption={Creating Solcore materials from the Sopra database.}, label={lst:sopra}]
# Normal GaAs material with all  the parameters
GaAs = material("GaAs")()
	
# Sopra version, with only optical constants
GaAs_sopra = material("GaAs", sopra=True)()
	
# Ni, Au and SiO2 are only in the Sopra database, so there is no need to include a flag
Ni = material("Ni")()
Au = material("Au")()
SiO2 = material("SiO2")()
	
# Creating materials with different alloy compositions
AlGaAs_sopra = material("AlGaAs", sopra=True)(Al=0.4)
HgCdTe = material("HgCdTe")(Cd=0.25)
	
# Relaxed SiGe alloys
SiGe25 = material("ReSiGe")(Si=0.25)
SiGe75 = material("ReSiGe")(Si=0.75)
\end{lstlisting}

\begin{figure*}
	\centering
	\includegraphics[width=\textwidth]{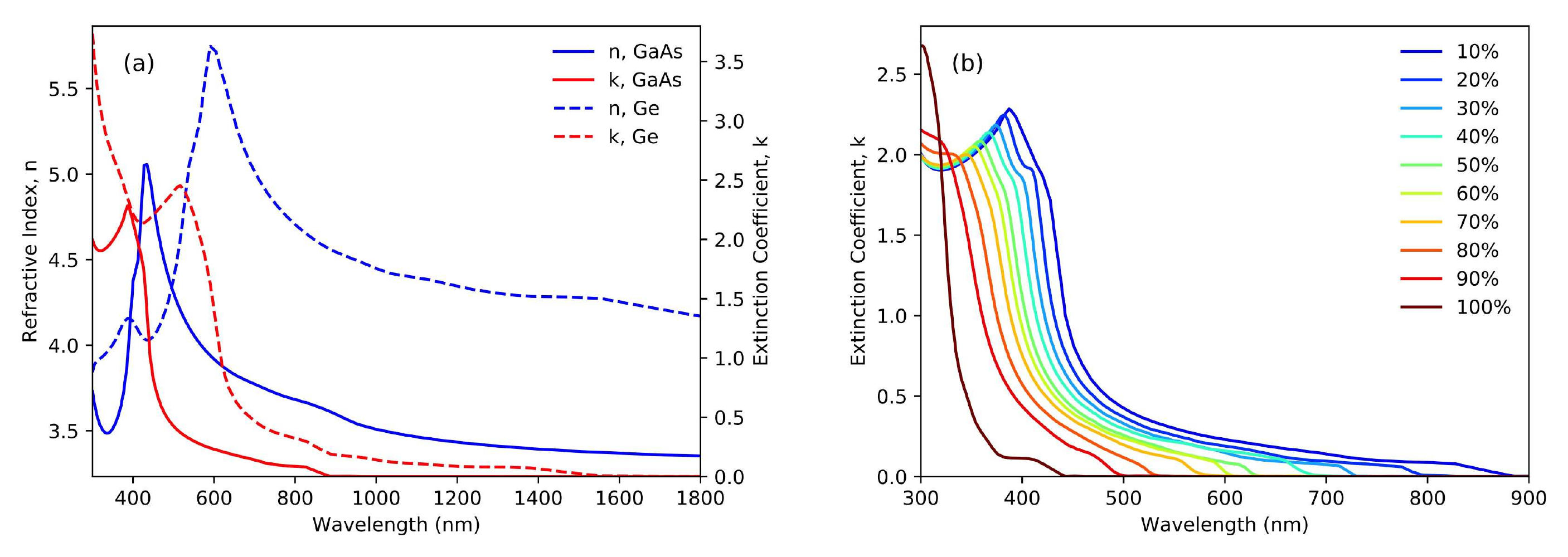}
	\caption{Example output from the Sopra S. A. optical constant database accessed from Solcore. (a) Refractive index and extinction coefficient data for GaAs (solid lines) and Ge (dashed lines). (b) Interpolated AlGaAs extinction coefficient data with aluminium fractions ranging from 10 to 100\%.}
	\label{fig:sopra}
\end{figure*}

Figure \ref{fig:sopra} highlights example output from the Sopra materials library with \ref{fig:sopra}a showing GaAs and Ge optical constant data and \ref{fig:sopra}b showing interpolated AlGaAs extinction coefficients for a range of aluminium fractions.   

\subsection{Quantum solver}

The electronic band structure of semiconductor materials is responsible for their light absorption and emission properties as well as for many of their transport properties, ultimately depending on the carriers' effective masses. These properties are not intrinsic to the material, but depend on external factors, too, most notably the strain and the quantum confinement. 

Given the crystalline nature of most semiconductor materials, there will be strain whenever two materials with different crystal lattice constants are grown on top of each other pseudomorphically. Even those with the same lattice constant might be under stress due to other effects such as the presence of impurities or if used at different temperatures having dissimilar thermal expansion coefficients. Quantum confinement, in turn, takes place when the size of the semiconductor material in one or more dimensions is reduced to a few nanometres. In that situation, the energy levels available to the carriers become quantized in the direction of confinement, also changing the density of states. Both conditions take place simultaneously when dealing with strain-balanced quantum wells (QW). 

Quantum wells - and more recently quantum wires - have been employed to tune the absorption properties of high efficiency solar cells for the past two decades. The need for appropriate tools to study them in the context of photovoltaics led to the development of the simulation models that were the seed of Solcore~\cite{Paxman:1993ej, Nelson:1997fb, Nelson:1999hm, Fuhrer:2011gw}. As strained materials with quantum confinement, special care must be taken to obtain a sensible set of parameters for the QW structures, including the band edges with confined energy levels, the effective masses and the absorption coefficient. 

Solcore's approach for evaluating the properties of QWs involves calculating first the effect of strain using a 8-band Pikus-Bir Hamiltonian (Section \ref{sec:8bands}), treating each material in the structure as bulk, and then using the shifted bands and effective masses to solve the 1D Schödinger equation, after a proper alignment between layers (Section \ref{sec:schrodinger})~\cite{Fuhrer:2014cz}. Finally, the absorption coefficient is calculated based on the 2D density of states, including the effect of excitons (Section \ref{sec:qw_abs}).  

\subsubsection{Bulk 8-band k\textbullet p calculator}\label{sec:8bands}
There are many numerical methods to calculate the band structure of a material with a varied degree of sophistication and complexity, such as the tight binding, pseudopotential, Green's function or \textbf{k$\cdot$ p} methods. Solcore includes a modified 8-band Pikus-Bir Hamiltonian to calculate the band structure of bulk materials under biaxial strain \cite{Tomic:2006dr}, considering the coupling between the conduction, heavy hole, light hole and split-off bands. The eigenfunctions $\Psi$ and eigenstates $E$ are the solutions of the following equation, where $\mathcal{H}$ is the Pikus-Bir hamiltonian:

\begin{strip}
\begin{equation}
\mathcal{H}\Psi = \left(
\begin{array}{cccccccc}
E_{cb} & -\sqrt{3}T & \sqrt{2}U	& -U	             & 0                 & 0                & -T^*           & -\sqrt{2}T^* \\
            & E_{hh}       & \sqrt{2}S	& -S	              & 0                & 0                & -R               & -\sqrt{2}R    \\
            &                   & E_{lh}       & -\sqrt{2}Q	& T^*	         & R                & 0                 & \sqrt{3}S    \\
            &				 &					& E_{so}       & \sqrt{2}T^*	& \sqrt{2}R	& -\sqrt{3}S & 0     \\
            &				&					&					& E_{cb}		& -\sqrt{3}T^* &\sqrt{2}U	& -U \\
            &				&					&					&					& E_{hh}		& \sqrt{2}S^*	& -S^* \\
            &				&					&					&					&					& E_{lh}			& -\sqrt{2}Q \\
            &				&					&					&					&					&					& E_{so} \\
\end{array}
\right)\Psi = E\Psi
\end{equation}
\end{strip}

Here, the sub-diagonal elements are just the complex conjugate of the corresponding upper elements. Diagonal terms have three components: the information about the unstrained band edges, a kinematic term, and a strain term. As an example, the term $E_{cb}$ is given by:

\begin{align}
E_{cb} &= E_{c0} + O_k + O_{\epsilon} \\
O_k & = \frac{\hbar^2}{2m_0} \gamma_c \left( k_x^2 + k_y^2 + k_z^2 \right) \\
O_{\epsilon} & = a_c \left( \epsilon_{xx} +   \epsilon_{zz} +  \epsilon_{zz} \right)
\end{align}

\noindent where $E_{c0}$ is the position of the unstrained conduction band edge, $\gamma_c$ a modified Luttinger parameter, the $k_{i}$ the momenta in the different directions of the reciprocal space, $a_c$ the conduction band hydrostatic deformation potential and the $\epsilon_{ij}$s the strain tensor components. Off-diagonal terms have similar expressions, not involving the unstrained band edges. A detailed description of all these terms and their origin can be found in Tomic~\cite{Tomic:2006dr}.

\begin{figure}
	\centering
  	\includegraphics[width=\columnwidth]{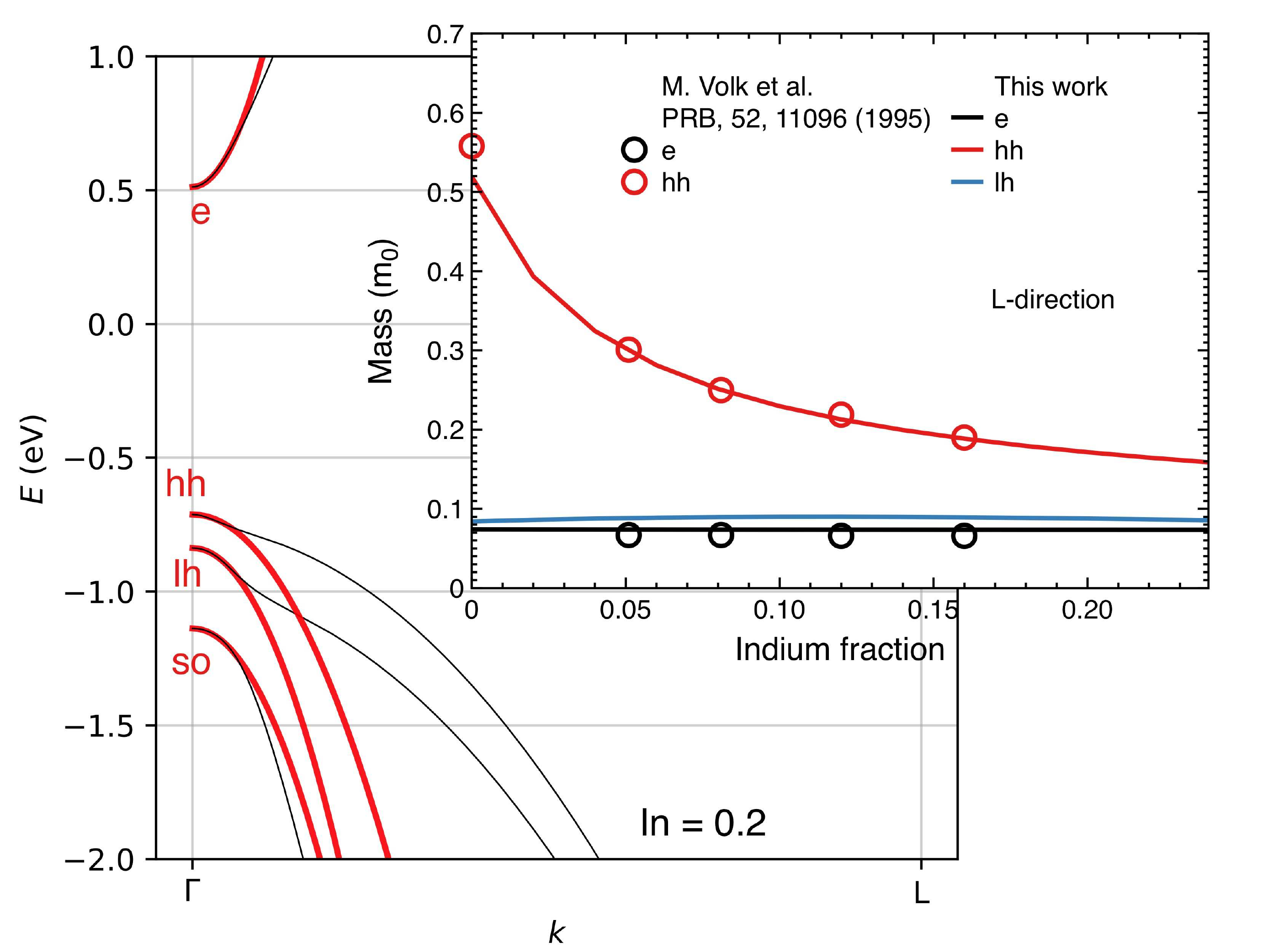}
  	\caption{Band structure of In$_{0.2}$Ga$_{0.8}$As calculated with the bulk \textbf{k$\cdot$ p} solver. The inset shows the effective mass determined for a range of indium fractions and a comparison with the experimental data from Volk et al.~\cite{Volk:1995iq}}
  	\label{fig:EffectiveMasses}
\end{figure}

This system is readily solved for the given $k_i$ using Numpy's \texttt{linalg.eig} function, which provides as output the eigenfunctions and the corresponding eigenvalues. Typically, we are interested in the new band edges due to the effect of strain and the resulting effective masses, given by the curvature of the bands near $k_i$ = 0. Fig. \ref{fig:EffectiveMasses} shows an example of the bands calculated in this way for the case of a strained InGaAs layer grown pseudomorphically on GaAs, and the resulting dependence of the effective mass with the indium content of the layer. Notice that, due to the effect of strain, the heavy and light hole bands are no longer degenerate at the gamma point $\Gamma$ ($k$ = 0).

\subsubsection{1D Schr\"odinger equation}\label{sec:schrodinger}
Once the new band edges and effective masses for each of the materials forming the quantum well structure are known, the quantum properties can be calculated by solving the 1-dimensional Schr\"odinger's equation. Solcore uses the method described by Frensley~\cite{Frensley:1992jj}, which allows calculation of the eigenvalues and eigenvectors of an arbitrary potential. However, only closed and periodic boundary conditions are implemented. Solcore does not incorporate the Quantum Transmitting Boundary Method (QTBM) described by Frensley, meaning that unbounded states will be, in general, not correct. A tridiagonal matrix is constructed by writing the variable effective mass Schr\"odinger's equation over a series of mesh points. The eigenvalues of the matrix correspond to the allowed energy levels of the system. Thus, the system of equations to solve over the mesh points is given by:

\begin{align}
\mathcal{H}\Psi_i = -s_i\Psi_{i-1} + d_i\Psi_i - s_{i+1}\Psi_{i+1} =  E\Psi_i
\end{align}

\noindent where $s_i$ and $d_i$ depend on the mesh spacing $\Delta$ and the position dependent potential $V_i$ and effective masses $m_i$ as:

\begin{align}
d_i &= \frac{\hbar^2}{4 \Delta^2 m_0}  \left(  \frac{1}{m_{i-1}} + \frac{2}{m_{i-1}} + \frac{1}{m_{i+1}}\right) + V_i \\
s_i &= \frac{\hbar^2}{4 \Delta^2 m_0}  \left(  \frac{1}{m_{i-1}} + \frac{1}{m_{i-1}}  \right)
\end{align}

This system is solved using the tools available in the Scipy package for solving sparse linear systems of equations, in particular \texttt{sparse.linalg.eigs}. 

\begin{figure*}
	\centering
  	\includegraphics[width=1\textwidth]{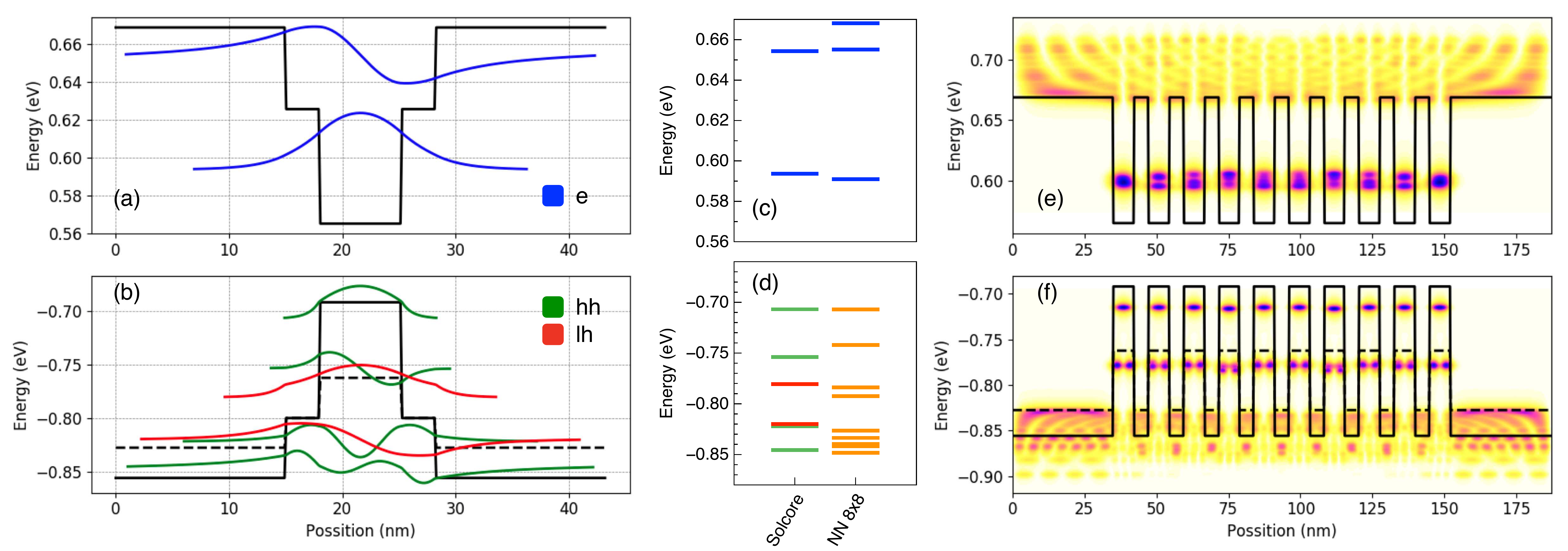}
  	\caption{(a), (b) Band profile of  a single 7.2 nm-thick In$_{0.15}$Ga$_{0.85}$As QW with 3 nm GaAs interlayers and GaAs$_{0.9}$P$_{0.1}$ barriers. (c), (d) comparison of the energy levels calculated by Solcore and using the more rigorous 1D 8-band kp solver implemented in Nextnano~\cite{Birner:2007gm}. (e), (f) 1D-LDOS of a multi-QW structure with 10 coupled QWs, as before but without interlayers. }
  	\label{fig:qw}
\end{figure*}

Fig. \ref{fig:qw} shows two examples of the band profile and wavefunctions calculated by Solcore. The first one (Fig. \ref{fig:qw}a and b) is a single InGaAs QW with GaAs interlayers and GaAsP barriers. The strain and quantum confinement shift the light hole valence band (dashed line) with respect to the heavy hole valence band (continuous line). In the GaAsP barriers, under tensile strain, this shift is in the opposite direction to the shift inside the QW, which is under compressive strain and experiences the effects of confinement. Fig. \ref{fig:qw}c and d compares the position of the energy levels predicted by Solcore in this structure with the more rigorous treatment of the 1-dimensional 8-band kp solver implemented by Nextnano++. As shown, the electron energy levels are barely modified, but hole levels are shifted due to the coupling between the bands~\cite{Birner:2007gm}. The in-plane dispersion when using a 8-band solver will no longer be parabolic and we would expect this to have an impact into the absorption coefficient and especially the selection rules for the transitions due to the band mixing effects (see Section~\ref{sec:qw_abs}).  

Finally, Fig. \ref{fig:qw}e and f shows 1D the local density of states (LDOS) of a a multi-QW structure with thin barriers, including a Lorentzian broadening of 5 meV. For the electrons, there is strong coupling between the QWs, resulting in a range of energies for the ground state. The heavy hole ground states are too deep, resulting in lower coupling between wells. This figure also shows the artefacts in the LDOS of unbound states due to the use of closed boundary conditions rather than QTBM. 

\subsection{Critical-Point Parabolic-Band Optical Constant Model}

Understanding the optical response of both established and novel materials is crucial to effective solar cell design. To efficiently model the complex dielectric function of a material Solcore incorporates an optical constant calculator based on the well-known Critical-Point Parabolic-Band (CPPB) formalism popularised by Adachi \cite{Adachi:1987gd, Adachi:1989km, Adachi:1989dl}. In this model, contributions to \textit{$\epsilon_2(\omega)$} from critical points in the Brillouin Zone at which the probability for optical transitions is large (van Hove singularities) are considered. The transition probability for such transitions is proportional to the joint density of states (JDOS) $\textbf{J}_{cv}(\omega)$, which describes the number of available electronic states between the valence and conduction bands at given photon energy. The imaginary part of the complex dielectric function is related to the JDOS by:

\begin{equation}
\label{eqn:JDOS}
\epsilon_2(\omega) = \frac{4 \hbar^2 e^2}{\pi \mu_{0}^{2} \omega^2} \left| \left \langle c | p | v \right \rangle \right|^2 \textbf{J}_{cv}(\omega)
\end{equation}

\noindent Where $\left| \left \langle c | p | v \right \rangle \right|$ is the momentum matrix element for transitions from the valence band ($v$) to the conduction band ($c$). Critical point transitions are considered at the following points of symmetry in the band structure: \textit{$E_0$} corresponds to the optical transition at the \textit{$\Gamma$} point and \textit{$E_0 + \Delta_0$} to the transition from the spin-orbit split off band to the conduction band at the \textit{$\Gamma$} point. \textit{$E_1$} and \textit{$E_1 + \Delta_1$} denote the transitions from the valence heavy-hole (HH) band and the valence light-hole (LH) band respectively to the conduction band at the \textit{L} point. The \textit{$E'_0$} triplet and \textit{$E_2$} transitions occur at higher energies, between the HH band and the split conduction bands at the $\Gamma$ point as well as across the wide gap \textit{X} valley. The model also includes contributions from the lowest energy indirect band-gap transition and the exciton absorption at the \textit{$E_0$} critical point. The contributions listed above are summed to compute the overall value of $\epsilon_2(\omega)$. The real and imaginary components of the overall complex dielectric function $\epsilon(\omega) = \epsilon_1(\omega) - i\epsilon_2(\omega)$ are then related via the Kramers-Kronig relations;

\begin{equation}
	\label{eqn:KKR_1}
	\epsilon_1(\omega) = 1 + \frac{2}{\pi} \int_{0}^{\infty} \frac{\omega' \epsilon_2(\omega')}{(\omega')^2 - \omega^2} d\omega'
\end{equation}

\begin{equation}
	\label{eqn:KKR_2}
	\epsilon_2(\omega) = - \frac{2}{\pi} \int_{0}^{\infty} \frac{\epsilon_1(\omega')}{(\omega')^2 - \omega^2} d\omega'
\end{equation}

The CPPB model included with Solcore also incorporates a modification to the critical point broadening present in Adachi's description, which is shown to produce a poor fit to experimental data in the vicinity of the $E_0$ and $E_1$ critical points \cite{Rakic:1996en}. To give a more accurate description of the broadening of the optical dielectric function, Kim et al. proposed that a frequency-dependent damping parameter be used to replace the damping constant given by Adachi at each critical point \cite{Kim:1992hh, Kim:1993bo};

\begin{equation}
	\label{eqn:Kim_damping}
	\Gamma'(\omega) = \Gamma exp \left[ -\alpha \left( \frac{\hbar \omega - E_0}{\Gamma}\right) ^2 \right]  
\end{equation}

\noindent Where $\Gamma$ is the damping constant used by Adachi and $\alpha$ describes the shape of the lineshape broadening with $\alpha = 0$ producing purely Lorentzian character and $\alpha = 0.3$ producing a good approximation to Gaussian broadening. \par

The Solcore module \verb|absorption_calculator| contains the CPPB model within the \verb|Custom_CPPB| class. The class offers a flexible way to build up the optical constant model by adding indivdual critical point contributions through the \textit{Oscillator} structure type within Solcore. In addition to the oscillator functions described by Adachi the \verb|Custom_CPPB| class also provides additional oscilator models and the Sellmeier equation for describing the real part of the dielectric function for non-absorbing materials \cite{Woollam:2012wp}. For example, the code in Listing \ref{lst:CPPB} calculates the complex dielectric function of GaAs.

\begin{strip}
\begin{lstlisting}[language=Python, caption={Modelling of the $n$ and $k$ based on the CPPB model.}, label={lst:CPPB}]
	from solcore.absorption_calculator.Custom_CPPB import Custom_CPPB
	import numpy as np

	# Generate a list of energies over which to calculate the model dielectric function.
	E = np.linspace(0.2, 5, 1000)
	
	# Class object is created, CPPB_Model
	CPPB_Model = Custom_CPPB()
	
	# The MatParams method loads in the desired material parameters as a dictionary variable.
	MatParams = CPPB_Model.Material_Params("GaAs")
	
	# The oscillator type and material parameters are both passed to individual oscilators in the
	# 'Oscillator' structure.
	Adachi_GaAs = Structure([
	    Oscillator(oscillator_type="E0andE0_d0", material_parameters=MatParams),
	    Oscillator(oscillator_type="E1andE1_d1", material_parameters=MatParams),
	    Oscillator(oscillator_type="E_ID", material_parameters=MatParams),
	    Oscillator(oscillator_type="E2", material_parameters=MatParams)
	])
	
	Output = CPPB_Model.eps_calc(Adachi_GaAs, E)
\end{lstlisting}
\end{strip}

\begin{figure}
	\centering
	\includegraphics[width=\columnwidth]{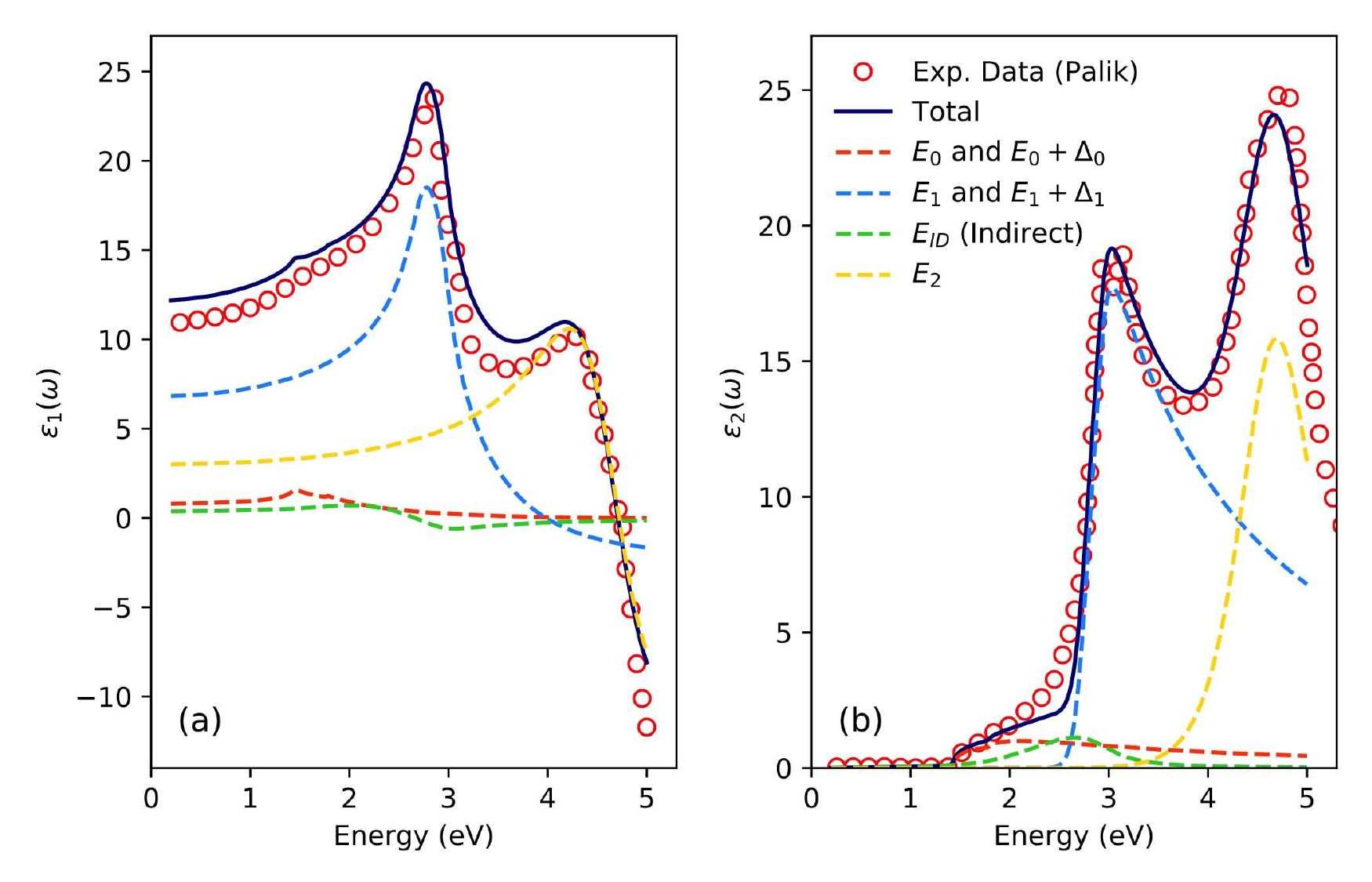}
	\caption{Output of the CPPB model provided by Solcore fit to existing experimental data for GaAs, from Palik \cite{Palik:1997ej}}
	\label{fig:cppb_eps}
\end{figure}

Figure \ref{fig:cppb_eps} shows the real and imaginary components of the complex dielectric function of GaAs as calculated by the \verb|Custom_CPPB| class. The model, using a set of parameters for GaAs similar to those specified in \cite{Adachi:1989km}, shows excellent agreement with the experimental data taken from Palik \cite{Palik:1997ej}. For a recent demonstration of Solcore's CPPB model, please refer to the discussion on Wilson \textit{et al.}~\cite{WilsonEUPVSEC2017}.

\subsection{QW absorption calculator}\label{sec:qw_abs}
For modelling the optical properties of QWs we use the method described by S. Chuang \cite{Chuang:1995vf}. The absorption coefficient at thermal equilibrium in a QW is given by:

\begin{equation}\label{eq:QW_abs2}
\begin{split}
\alpha_0(E) & = C_0(E) \sum_{n,m} |I_{hm}^{en}|^2 | \hat{e} \cdot \vec{p} |^2 \rho_{rmn}^{2D} \\ 
& \times \left[ H(E-E^{en} + E_{hm}) + F_{nm}(E) \right]
\end{split}
\end{equation}

\noindent where $|I_{hm}^{en}|^2$ is the overlap integral between the holes in level $m$ and the electrons in level $n$; $H$ is a step function, $H(x)$ = 1 for $x>0$, 0 and 0 for $x<0$, $\rho_{rmn}^{2D}$ is the 2D joint density of states, $C_0$ a proportionality constant dependent on the energy, and $F$ the excitonic contribution, which will be discussed later. 

\begin{align}\label{eq:qw_abs}
C_0 (E) & =    \frac{\pi q^2 \hbar }{n_r c \epsilon_0 m_0^2 E} \\
\rho_r^{2D} &= \frac{m_{rmn}^*}{\pi \hbar L}
\end{align}

Here, $n_r$ is the refractive index of the material, $m_{rmn} = m_{en} m_{hm} / (m_{en} + m_{hm})$ the reduced, in-plane, effective mass and $L$ an effective period of the quantum wells. The in-plane effective mass of each type of carriers is calculated for each level, accounting for the spread of the wavefunction into the barriers as~\cite{barnham_vvedensky_2001}:

\begin{align}\label{eq:in_plane}
m_{\perp} =  \int_{0}^{L} m(z) | \psi(z) |^2
\end{align}

This in-plane effective mass is also used to calculate the local density of states shown in Figure~\ref{fig:qw}b. In Eq.~\ref{eq:QW_abs2}, $| \hat{e} \cdot \vec{p} |^2$ is the momentum matrix element, which depends on the polarization of the light and on the Kane's energy $E_p$, specific to each material and determined experimentally. For band edge absorption, where $k$ = 0, the matrix elements for the absorption of TE and TM polarized light for the transitions involving the conduction band and the heavy and light holes bands are given in Table \ref{tab:matrix_elements}. As can be deduced from this table, transitions involving heavy holes cannot absorb TM polarised light. 

\begin{table}
\begin{center}
\begin{tabular}{c | c c }
 \hline
  				& TE 			& TM \\
 \hline
$c-hh$ 	& $3/2 M_b^2$ 		& 0\\ 
 \hline
$c-lh$ 		& $1/2 M_b^2$ 		& $2 M_b^2$ \\
 \hline
 \end{tabular}
 \end{center}
\caption{Momentum matrix elements for transitions in QWs. $M_b^2=m_0 E_p /6$ is the bulk matrix element.}
\label{tab:matrix_elements}
\end{table}

In addition to the band-to-band transitions, QWs usually have strong excitonic absorption, included in Eq. \ref{eq:qw_abs} in the term $F_{nm}$. This term is a Lorenzian (or Gaussian) defined by an energy $E_{nmx, j}$ and oscillator strength $f_{ex, j}$. It is zero except for $m=n \equiv j$ where it is given by Klipstein et al.~\cite{Klipstein:2000gu}:

\begin{align}
F_{nm} &= f_{ex, j} \mathcal{L}(E - E_{nmx, j}, \sigma) \\  
E_{nmx, j} &= E^{en} - E_{hm} - \frac{R}{(j-\nu)^2} \\
f_{ex, j} &= \frac{2R}{(j-\nu)^3} \\
R &= \frac{m_r q^4}{2 ( 4\pi \epsilon_r \epsilon_0)^2 \hbar^2 }
\end{align}

Here, $\nu$ is a constant with a value between 0 and 0.5 and $\sigma$ is the width of the Lorentzian, both often adjusted to fit some experimental data. In Solcore, they have default values of $\nu$ = 0.15 and $\sigma$ = 6 meV. $R$ is the exciton Rydberg energy~\cite{Chuang:1995vf}.

\begin{figure}
  	\centering
  	\includegraphics[width=\columnwidth]{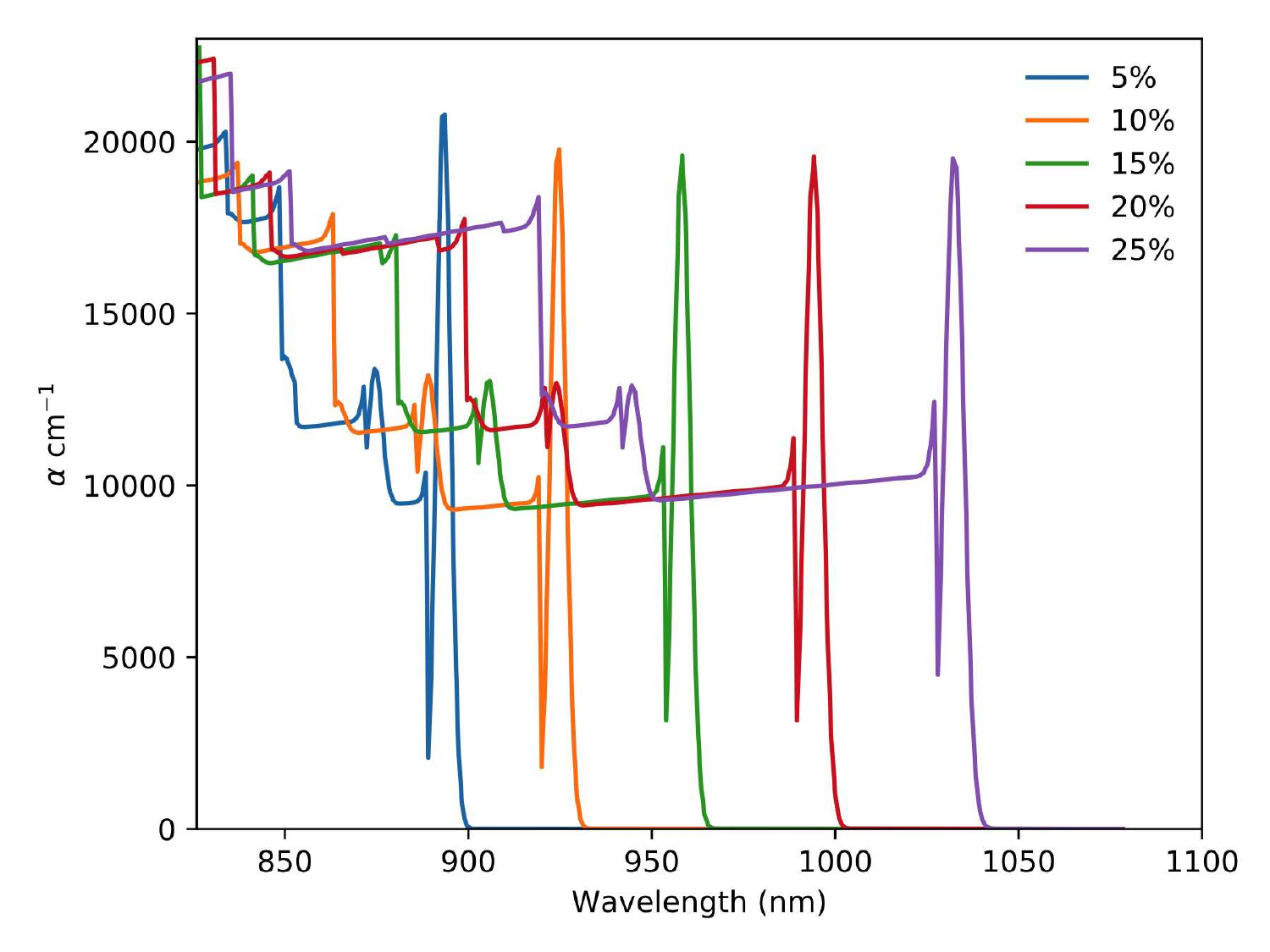}
  	\caption{Absorption coefficient of a single 7.2 nm-thick InGaAs QW with 3 nm GaAs interlayers and GaAs$_{0.9}$P$_{0.1}$ barriers as a function of the indium content.}
  	\label{fig:QW_absorption}
\end{figure}

Fig. \ref{fig:QW_absorption} shows the absorption coefficient of a range of InGaAs/GaAsP QWs with a GaAs interlayer and different In content. Higher indium content increases the depth of the well, allowing the absorption of less energetic light and more transitions. 

\section{Light sources} \label{sec:light sources}

Transforming sunlight into electricity is the final goal of any solar cell and it is therefore necessary to have a convenient way of creating, manipulating and modifying the properties of the spectrum of the light. Ideally, solar cells will be designed and evaluated under a standard solar spectrum - e.g. the air mass 1.5 direct solar spectrum, AM1.5D - but practical light sources are not standard. More often than not, we are interested in modelling the performance of a solar cell under the experimental spectrum of a solar simulator or lamp in a laboratory, simulated data calculated from atmospheric conditions (temperature, humidity, aerosol content, etc.) or even under real irradiance data measured at different locations worldwide. This can then be compared with the experimental performance or tailored to work best under certain conditions.  

The Solcore module \texttt{light\_source} is designed to deal easily with different light sources. It has direct support for:

\begin{itemize}
	\item Gaussian emission, typical of lasers and light emitting diodes.
	\item Black-body radiation, characteristic of halogen lamps defined by a temperature, but also used very often to simulate the spectrum of the Sun, very close to a black body source at 5800 K. 
	\item Standard solar spectra: the extraterrestial spectrum AM0 and the two terrestial ones, AM1.5D and AM1.5G as defined by the ASTM G173 - 03(2008) standard.
	\item Irradiance models, using location, time and atmospheric parameters to calculate a synthetic solar spectrum. Solcore includes two models: SPECTRAL2, fully implemented in Python, and an interface to SMARTS binaries (which need to be installed separately), which greatly simplifies its use in batch mode. 
	\item User-defined irradiances, provided externally from a database or any other source, allowing for maximum flexibility. 
\end{itemize}

\begin{figure}
  	\centering
  	\includegraphics[width=\columnwidth]{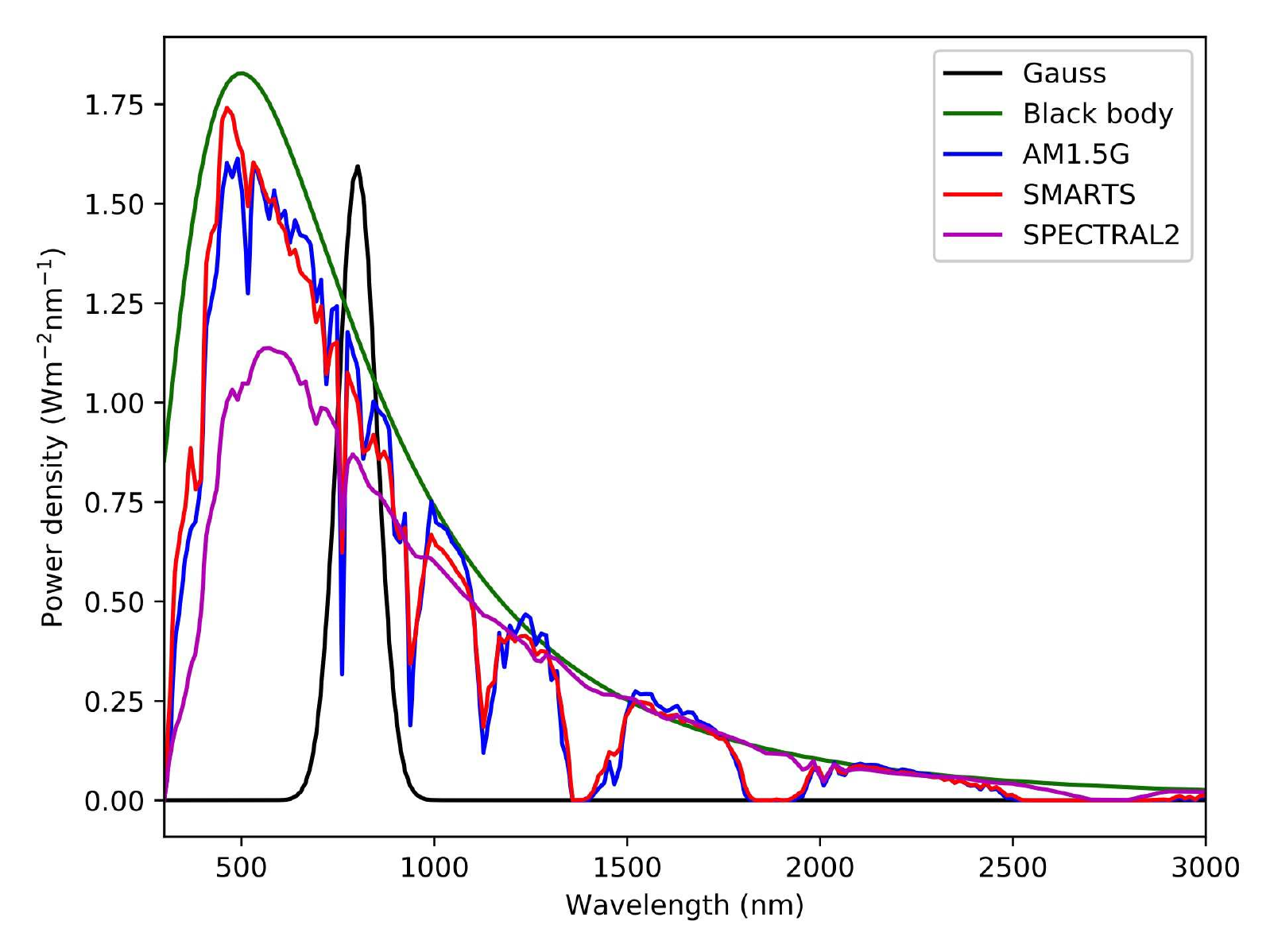}
  	\caption{Plot of the spectra of different light sources.}
  	\label{fig:light_source}
\end{figure}

The syntax in all cases is simple and intuitive considering the type of source that needs to be created. In the case of the irradiance models, which often have a large number of inputs, Solcore defines a set of default values, so only those that are different need to be provided. The code in Listing \ref{lst:light_sources} illustrates the creation of several light sources using the minimum required input in each case. A plot of those light sources is shown in Figure \ref{fig:light_source}.

\begin{strip}
\begin{lstlisting}[language=Python, caption={Example of the use of the LightSource class.}, label={lst:light_sources}]
import numpy as np
from solcore.light_source import LightSource

#The wavelength range of the spectra
wl = np.linspace(300,3000,200)	

gauss = LightSource(source_type='laser',x=wl,center=800,linewidth=50,power=200)
bb = LightSource(source_type='black body',x=wl,T=5800,entendue='Sun')
am15g = LightSource(source_type='standard',x=wl,version='AM1.5g')
smarts = LightSource(source_type='SMARTS',x=wl)
spectral = LightSource(source_type='SPECTRAL2',x=wl)
\end{lstlisting}
\end{strip}

Once created, specific parameters of the light sources can be easily modified without the need for creating the source from scratch. That is particularly useful for the irradiance models, where we might be interested in getting the spectrum as a function of a certain parameter (e.g. the hour of the day, or the humidity) without changing the others. For example, \verb+smarts.spectrum(HOUR=11)+ and \verb+smarts.spectrum(HOUR=17)+ will provide the spectrum of the SMART light source defined above calculated at 11h and at 17h, respectively; all additional parameters have the default values. This method has been used to model experimental solar irradiances measured by different spectroradiometers based on the local atmospheric conditions~\cite{Galeano:2016to}.

A final, very convenient feature of the LightSource class is the ability to request the spectrum in a range of different units. The default is power density per nanometer, but other common units are power density per eV or photon flux per nanometer, among others. While these unit conversions are straightforward, it is often an initial source of errors due to missing constants or incompatible magnitudes.

The \texttt{light\_source} module has been described in the context of the solar spectrum, but it can be applied broadly where there is spectral data involved, such as the fitting of photoluminescence, electroluminescence or Raman spectra.

\section{Optical solvers} \label{sec:Optical solvers}

The purpose of the optical solvers is to obtain the fraction of incoming light reflected, absorbed and transmitted in a solar cell as a function of the wavelength of the light and the position inside the structure. Solcore includes three models to tackle this problem: Beer-Lambert law (BL), transfer matrix method (TMM) and rigorous coupled wave analysis (RCWA). At the moment, Solcore does not have explicit support for light trapping effects using general textured surfaces, which are usually present in silicon solar cells. However, this can be implemented to a large extent using the RCWA method, although not very efficiently. Additionally, the reflected, absorbed and transmitted light can be calculated externally and then provided as input to Solcore to obtain the electrical properties of a solar cell structure, giving it full flexibility. 

All the optical solvers apply to the solar cell structure as a whole, providing as output the fraction of light reflected ($R(\lambda)$), transmitted ($T(\lambda)$) and absorbed per unit length at a depth $z$ from the front surface ($A(\lambda, z)$).

\begin{figure}
	\centering
	\includegraphics[width=\columnwidth]{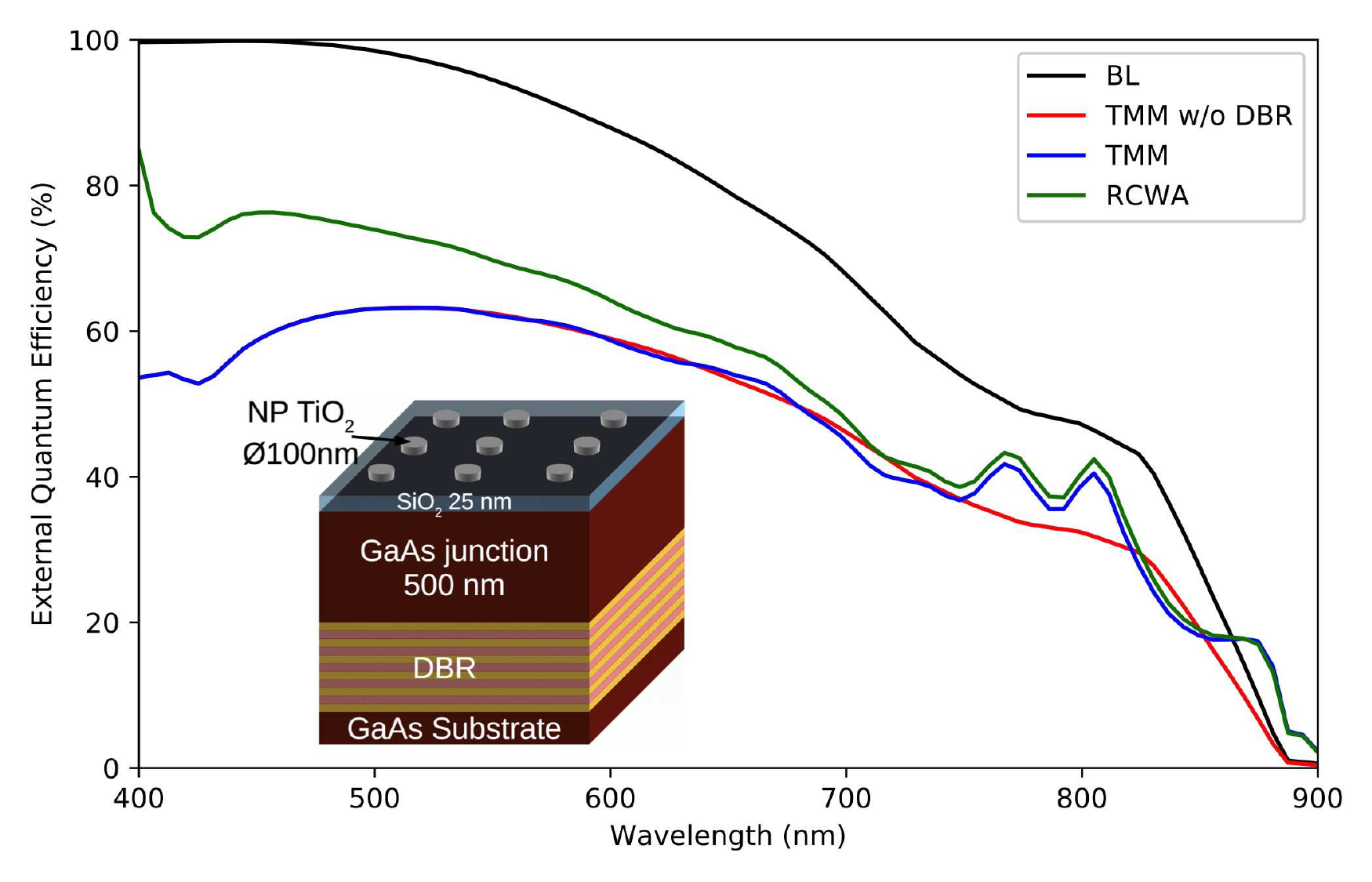}
	\caption{Comparison of the quantum efficiency of a thin GaAs solar cell with a distributed Bragg reflector (DBR) and an array of TiO2 nanoparticles (NP) on top calculated using the three optical solvers. The BL model ignores the NP and the DBR and does not include the front surface reflection, overestimating the result at all wavelengths. TMM correctly accounts for the reflection and the DBR but cannot model the effect of diffraction due to the NP layer. RCWA takes into account scattering from the NP, although it is significantly more time consuming. }
	\label{fig:optic_comparison}
\end{figure}

Figure~\ref{fig:optic_comparison} shows a comparison of the quantum efficiency of a thin GaAs solar cell with a distributed Bragg reflector (DBR) and an array of TiO$_2$ nanoparticles (NP) on top calculated using the three optical solvers described in the following sections. Supplementary Information shows the full code needed to produce these curves~\cite{alonso:sup_inf}. 

\subsection{Beer-Lambert law (BL)}

This is the simplest model to calculate the absorption in a multi-layer structure. It ignores all reflection at the interfaces - the front surface reflection can be provided externally, and is zero otherwise at all wavelengths - and the absorption per unit length as a function of the wavelength $\lambda$ and the position $z$ in layer \textit{n} is given by: 

\begin{equation} \label{eq:BL}
A_n(\lambda , z) =  \alpha_n (\lambda ) \exp \left( - \sum_{i=1}^{n-1} \alpha_i (\lambda ) w_i - \alpha_n (\lambda ) (z - z_n) \right) 
\end{equation}

\noindent where $\alpha_n$ is the absorption coefficient of layer \textit{n}, $w_n$ its thickness and $z_n$ the position of the beginning of the layer. Due to its simplicity, the BL law is used widely in photovoltaics but in reality it is only applicable when the contrast in the refractive index between layers can be ignored and when there is strong absorption, reducing the effects of light reflection at the interfaces.

\subsection{Transfer matrix method (TMM)}

\noindent In order to evaluate the realistic optical behaviour of a solar cell design it is important to consider the interaction of incident electromagnetic (EM) radiation with a succession of both absorbing and non-absorbing planar layers. The combined optical response of such a layered structure is crucial when considering the minimisation of extrinsic front surface reflection losses \cite{Wilson:2016jv}, the emissivity in the mid-IR of low emissivity coatings for hybrid PV-thermal applications ~\cite{AlonsoAlvarez:RlDRyvqY} and also when studying the optical constants and layer thicknesses of material using the experimental technique of spectroscopic ellipsometry. Therefore, Solcore evaluates the interaction of incident EM radiation through a layered structure using the TMM. The incident light radiation takes the form of homogeneous, electromagnetic plane-polarised waves and is represented by components describing the electric, \textbf{\textit{E}} and magnetic field strengths, \textbf{\textit{H}}:

\begin{align}
	\label{Eqn_E}
	\mathbf{E} &= \mathcal{E} exp \left[ i \omega t - \left( \frac{2 \pi N}{\lambda} \right) z + \varphi \right] \\
	\label{Eqn_H}
	\mathbf{H} &= \mathcal{H} exp \left[ i \omega t - \left( \frac{2 \pi N}{\lambda} \right) z + \varphi^{'} \right]
\end{align}

\noindent Where $\mathcal{E}$ and $\mathcal{H}$ denote the electric and magnetic field amplitudes respectively, \textit{N} the complex refractive index, \textit{z} the distance in the direction of propagation, $\omega$ the angular frequency and $\lambda$ the wavelength of radiation. $\varphi$ and $\varphi^{'}$ represent arbitrary phase angles for both the electric and magnetic travelling wave components and are not independent of each other. The characteristic transfer matrix for evaluating the interaction between planar electric and magnetic waves at the interface of \textit{n} thin films on a semi-infinite substrate is derived in detail elsewhere \cite{Macleod:2001ti} and given by the equation:

\begin{equation}
	\label{eqn:tmm}
	\begin{bmatrix} 
	\mathbf{E} \\
	\mathbf{H}
	\end{bmatrix}
	=				
	\left\{
	\prod_{q=1}^{n}
	\begin{bmatrix}
	cos(\delta_q) & [i sin(\delta_q)]/ \eta_q \\ 
	i \eta_q sin(\delta_q) & cos(\delta_q) 
	\end{bmatrix}
	\right\}
	\begin{bmatrix}
	1 \\
	\eta_m
	\end{bmatrix}
\end{equation}  

\noindent Where $\delta_q$ is defined as the phase factor of the $q_{th}$ planar layer, $\eta_q$ the optical admittance of the $q_{th}$ layer and $\eta_m$ the optical admittance of the substrate. The layer closest to the incident medium is evaluated first before working through the \textit{n} layer structure in order. The term $\delta_q$ describes the phase shift required to translate the \textit{z} coordinate of the \textit{E} and \textit{H} interactions by the thickness of each layer, \textit{q}. The spectrally varying Fresnel coefficients describing reflection, transmission and absorption of the multi-layer structure can be calculated from the solutions to equation \ref{eqn:tmm} at discrete wavelengths:

\begin{align}
\label{Eqn_Re}
R &= \left(\frac{\eta_0 \mathbf{E} - \mathbf{H}}{\eta_0 \mathbf{E} + \mathbf{H}}\right)\left(\frac{\eta_0 \mathbf{E} - \mathbf{H}}{\eta_0 \mathbf{E} + \mathbf{H}}\right)^* \\
\label{Eqn_Tr}
T &= \frac{4 \eta_0 Re(\eta_m)}{(\eta_0 \mathbf{E} + \mathbf{H})(\eta_0 \mathbf{E} + \mathbf{H})^*} \\
\label{Eqn_Ab}
A &= \frac{4 \eta_0 Re(\mathbf{E}\mathbf{H}^* - \eta_m)}{(\eta_0 \mathbf{E} + \mathbf{H})(\eta_0 \mathbf{E} + \mathbf{H})^*}
\end{align}

The implementation of the TMM in Solcore uses the freely available \verb|tmm| module developed by Byrnes \cite{Byrnes:lxGXf7Pu}. The multi-layer optical stack is built up using Solcore's \textit{Structure} object. Some example code evaluating the TMM for a triple layer anti-reflection coating (ARC) on top of conventional multi-junction solar cell materials AlInP and GaInP is included in Listing \ref{lst:RAT}.

\begin{strip}
\begin{lstlisting}[language=Python, caption={Calculation of the reflection, absorption and transmission of a structure.}, label={lst:RAT}]
	# The optical stack is built defining layer thickness, wavelength range and material 
	# n and k data.
	OptiStack = Structure([
	    [117, 1240/E_eV, mgf_nk[1], mgf_nk[2]],
	    [80, 1240/E_eV, sic_nk[1], sic_nk[2]],
	    [61, 1240/E_eV, zns_nk[1], zns_nk[2]],
	    [25, 1240/E_eV, alinp_nk[1], alinp_nk[2]],
	    [350000, 1240/E_eV, gainp_nk[1], gainp_nk[2]]
	  ])
	
	# The Reflection, Transmission and Absorption is evaluated for a range of incident 
	# angles (in degrees).
	angles = np.linspace(0, 80, 10)
	RAT_angles = []
	
	for theta in angles:
	
	    rat_data = []
	    # Calculate RAT data...
	    rat_data = calculate_rat(OptiStack, angle=theta, wavelength=1240 / E_eV)
	
	    RAT_angles.append((theta, rat_data["R"], rat_data["A"], rat_data["T]))
\end{lstlisting}
\end{strip}

\begin{figure*}
	\centering
	\includegraphics[width=\textwidth]{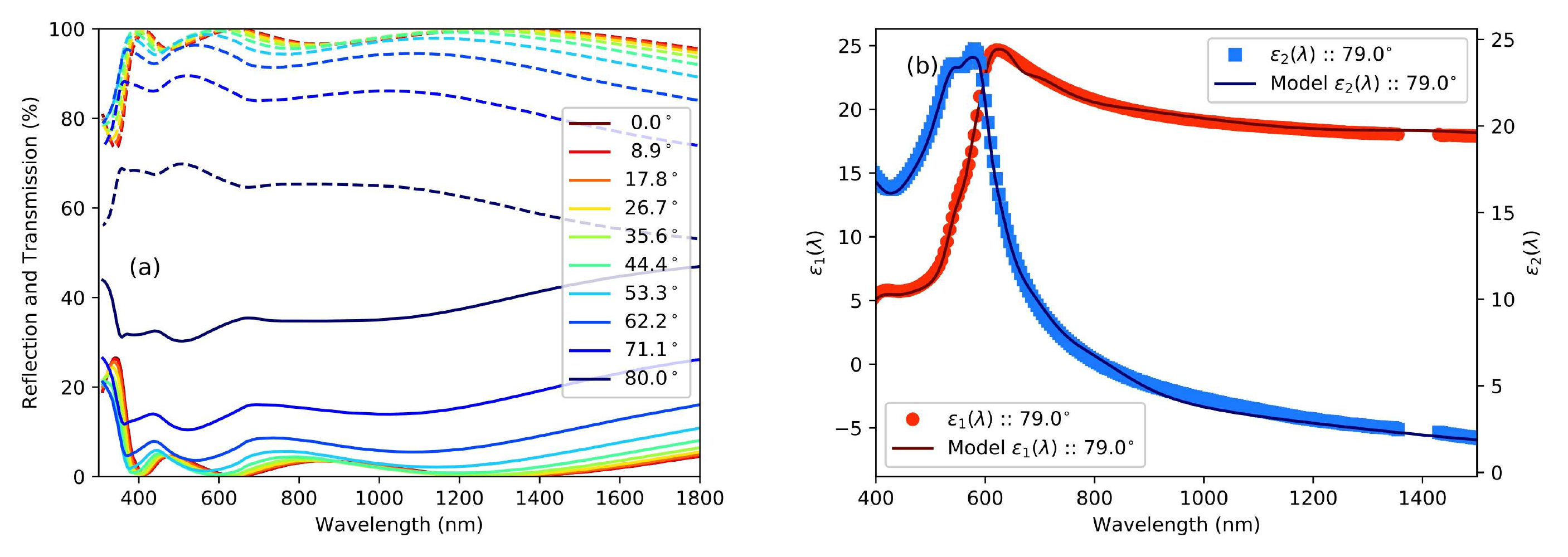}
	\caption{Example solutions from the TMM module in Solcore calculating Reflection, Transmission and complex dielectric function. (a) Spectral reflection (solid lines) and transmission (dashed lines) over a range of angles for the optimised triple layer ARC reported in \cite{Wilson:2016jv}. (b) The calculated real and imaginary parts of the complex dielectric function obtained from spectroscopic ellipsometry of a Ge substrate at $79^{\circ}$ (points), and the dielectric constant calculated using TMM for a model consisting of a semi-infinite Ge substrate and 4.5 nm GeO$_2$ (lines).}
	\label{fig:tmm_output}
\end{figure*}

\noindent Figure \ref{fig:tmm_output}a depicts calculated reflection and transmission from the solutions to the characteristic TMM equation over a range of incident angles for an optimised triple layer ARC design reported in \cite{Wilson:2016jv}. The solid lines indicate the optical reflection at the front surface whilst the dashed lines correspond to the transmitted light into the substrate, in this case taken to be the optically thick top sub-cell material of GaInP. \par

In addition to using the TMM for calculating the reflection, transmission and absorption in a multi-layer optical stack it can also be applied to the popular spectroscopic technique of ellipsometry. This measures a change in the polarisation of incident light reflected at the surface of a sample. A more detailed description of ellipsometry and its uses can be found elsewhere \cite{Woollam:2012wp}. The measured values are expressed as the angles Psi ($\Psi$) and Delta ($\Delta$), which are related to $\mathbf{R_s}$ and $\mathbf{R_p}$ (the Fresnel reflection coefficients for \textit{s} and \textit{p}-polarized light, respectively) by:\footnote{The definition of the phase $\Delta$ in the definition of $\rho$ (Eq. \ref{eqn:ellipsometry}), which affects the sign of $\epsilon_2$ in Eq. \ref{eqn:eps_ellips}, is the same here as in \cite{Woollam:2012wp}. However, other conventions are in use; for instance, $\Delta$ is defined with the opposite sign in \cite{Byrnes:lxGXf7Pu}.}

\begin{equation}
	\label{eqn:ellipsometry}
	\rho = \frac{\mathbf{R_p}}{\mathbf{R_s}} = tan(\Psi) e^{i \Delta}
\end{equation}

As the ratio between $\mathbf{R_s}$ and $\mathbf{R_p}$ is a complex quantity, phase change information is contained within $\Delta$ in Eq. \ref{eqn:ellipsometry}. The change in phase of the reflected electric and magnetic plane waves can be evaluated from the solution of Eq. \ref{eqn:tmm} and is given by:

\begin{equation}
\label{eqn:phase}
\varphi = arctan\left(\frac{Im[\eta_m (\mathbf{E}\mathbf{H}^* - \mathbf{H}\mathbf{E}^*)]}{(\eta_{m}^{2} \mathbf{E}\mathbf{E}^* - \mathbf{H}\mathbf{H}^*)} \right) 
\end{equation}

The complex dielectric function of a sample can be calculated from the experimental ellipsometry results with the solutions to Eq. \ref{eqn:ellipsometry}:

\begin{equation}
	\label{eqn:eps_ellips}
	\left\langle \epsilon \right\rangle = \epsilon_1 - i\epsilon_2 = sin^2(\theta) \left[ 1 + tan^2(\theta) \left( \frac{1 - \rho}{1 + \rho}\right)^2  \right] 
\end{equation}

Some example code, calculating the complex dielectric function from the ellipsometric response of a sample of Ge substrate is shown in Listing \ref{lst:ellipsometry}. The output of the model is shown in Figure \ref{fig:tmm_output}b and is compared with experimentally obtained data at an incident angle of $79 ^{\circ}$. Good agreement with the experimental data is observed when a thin 4.5 nm layer of germanium oxide (GeO$_2$) is included in the layer model. 

\begin{strip}
\begin{lstlisting}[language=Python, caption={Example of the calculation of the ellipsometric $\Psi$ and $\Delta$.}, label={lst:ellipsometry}]
from solcore.absorption_calculator import calculate_ellipsometry
from solcore.structure import Structure
	
# Input array of energies (in eV).
E_eV = np.linspace(0.7, 4.2, 1000)

Define the optical stack structure, a piece of Ge substrate with a thin Ge oxide layer.
# Layer 1 :: GeO2 native oxide layer
# Substrate :: Bulk Ge
OptiStack = Structure([
	    [4.5, 1240/E_eV, GeO2_nk["n"], GeO2_nk["k"]],  
	    [350000, 1240/E_eV, Ge_nk["n"], Ge_nk["k"]]      
		])
	
# Calculate ellipsometry variables, Psi and Delta
Exp_Angles = [75, 77, 79]
Out = calculate_ellipsometry(OptiStack, 1240/E_eV, angle=Exp_Angles)

# From calculated ellipsometry variables, Psi and Delta, the complex dielectric function 
# is computed using equations 25 and 27.
rho = lambda psi, delta: np.tan(psi) * np.exp(1j * delta)
eps = lambda r, theta: np.sin(theta)**2 * (1 + np.tan(theta)**2 * ((1 - r)/(1 + r))**2)
	
# Modelled data...
Mod_rho = rho(np.radians(Out["psi"][:,i]), np.radians(Out["Delta"][:,i]))
Mod_eps = eps(Mod_rho, np.radians(Exp_Angles[i]))
\end{lstlisting}
\end{strip}

\subsection{Rigorous coupled-wave analysis (RCWA)}

Finally, Solcore includes an interface to the $S^4$ solver (which must be installed separately), developed at Stanford University, in order to model solar cells with nanophotonic designs~\cite{Liu:2012gca}. $S^4$ is an implementation of RCWA, also sometimes referred to as the Fourier Model Method (FMM), which solves the linear Maxwell's equations in structures containing 2D periodicity. Structures with 2D periodicity can be found in advanced solar cell designs aiming, for example, to reduce the solar cell thickness by scattering the incoming light using a periodic diffraction grating at the front or rear of the absorbing layer \cite{Hylton2013, Mellor2017}.

$S^4$ defines structures by creating a layer stack of the desired materials using Solcore's \textit{Layer} and \textit{Junction} classes, in which each layer's composition can be modified by adding circles, rectangles, ellipses or a generalised polygon made of a specified Solcore material. Each layer is assumed to be infinitely periodic in the $x$ and $y$ direction, and uniform in the $z$ direction. A unit cell must be defined for the whole structure (using the \texttt{size} attribute in the user options), and each shape is placed at a specified location (and, where relevant, angular orientation) in the unit cell; as many shapes as necessary can be added, supplied as a list of dictionaries with the relevant parameters for each shape. An example for each type of shape supported by $S^4$ is shown in Listing \ref{lst:geometry_example}. The size of the unit cell in the $x$ and $y$ directions and the base size of the unit cell must be given in nm. The number of Fourier components used in the calculation \cite{Liu:2012gca} must also be specified; this is done using the \texttt{orders} attribute in the user options passed to the solver of choice.

\begin{strip}
\begin{lstlisting}[language=Python, caption={Creating geometry objects for use with the $S^4$ RCWA solver, which are added to Solcore \texttt{Layer} objects using the \texttt{geometry} attribute.}, label={lst:geometry_example}]
from solcore import si, material
from solcore.structure import Layer

Air = material('Air')(T=298)
TiO2 = material('TiO2', sopra=True)(T=298)  

# Define four geometry types, as a list of dictionaries: circles, rectanges, a general polygon 
# (in this case, a triangle) and ellipses. All the shapes are made of TiO2.
geometry_circles = [{'type': 'circle', 'mat': TiO2, 'center': (100, 100), 'radius': 50}]
geometry_rectangles = [{'type': 'rectangle', 'mat': TiO2, 'center': (200, 200), 'angle': 30, 
	'halfwidths': (50, 75)}] # rotation angle in degrees
geometry_polygons = [{'type': 'rectangle', 'mat': TiO2, 'center': (200, 200), 'angle': 0, 
	'vertices': ((-100, -100), (100, -100), (0, 100))}]		
# vertices must be defined counter-clockwise around the center (origin) for the unrotated polygon
geometry_ellipses = [{'type': 'rectangle', 'mat': TiO2, 'center': (200, 200), 'angle': 0, 'halfwidths': (50, 75)}]

# Define a layer of air with TiO2 nanopillars 50nm high.
cylinder_layer = Layer(width=si('50nm'), material=Air, geometry=geometry_circles)
	
\end{lstlisting}
\end{strip}

\section{Single-junction solar cells} \label{sec:electrical solvers}

Solcore includes four solvers to calculate the electrical properties of a single-junction device. In order of increasing accuracy, these are: detailed balance, 2-diode equation, depletion approximation and Poisson-drift-diffusion.

\subsection{Detailed balance (DB)}

This solver calculates the electrical properties of the junction by balancing the elementary processes taking place in the solar cell, carrier generation and radiative recombination, using the formalism described by Araújo and Martí~\cite{Marti:1996bh}. The method is widely used by the photovoltaic community to calculate the limiting conversion efficiencies of the different solar cell architectures or materials. The simplest DB formulation only needs an absorption edge energy and an absorptivity value above that edge. Out of this, the carrier generation and radiative recombination are calculated for different internal chemical potentials, equal to the external electrical bias, in the ideal case. Solcore includes this basic model, but also allows the user to provide a more complex absorption profile.   

The radiative recombination or thermal generation current $J_{rad}$ from the solar cell is calculated following the formalism described by Nelson et al.~\cite{Nelson:1997fb}, considering all the possible paths of the light absorbed by the cell and the reciprocity relation between emission and absorption. The total radiative current, using the generalized Planck equation, is given by:

\begin{equation}\label{eq:DB_radiative}
\begin{split}
J_{rad}(V, T) &= q \frac{2 n^2}{h^3c^2}  \int_{0}^{\infty } \frac{E^2}{e^{ \frac{E - qV}{k_bT}} - 1 } \\
& \times  \left[ \int_{S} A(E, \theta, \vec{s}) d\Omega \vec{dS} \right] dE \\
&=q \frac{2 n^2}{h^3c^2}  \int_{0}^{\infty } \frac{E^2}{e^{ \frac{E - qV}{k_bT}} - 1 } \\
& \times \left[ A_{front}(E)  + A_{back}(E)  \right] dE
\end{split}
\end{equation}

\noindent where A(E, $\theta$, $\vec{s}$) is the probability that a photon of energy $E$ will be emitted (absorbed) from the point $\vec{s}$ on the surface at an \textit{internal} angle $\theta$, and $A_{front}(E)$ and $A_{back}(E)$ are the combined probability of the photon to be emitted (absorbed) by the front and the back of the cell, respectively. 

\begin{figure}
  	\centering
  	\includegraphics[width=\columnwidth]{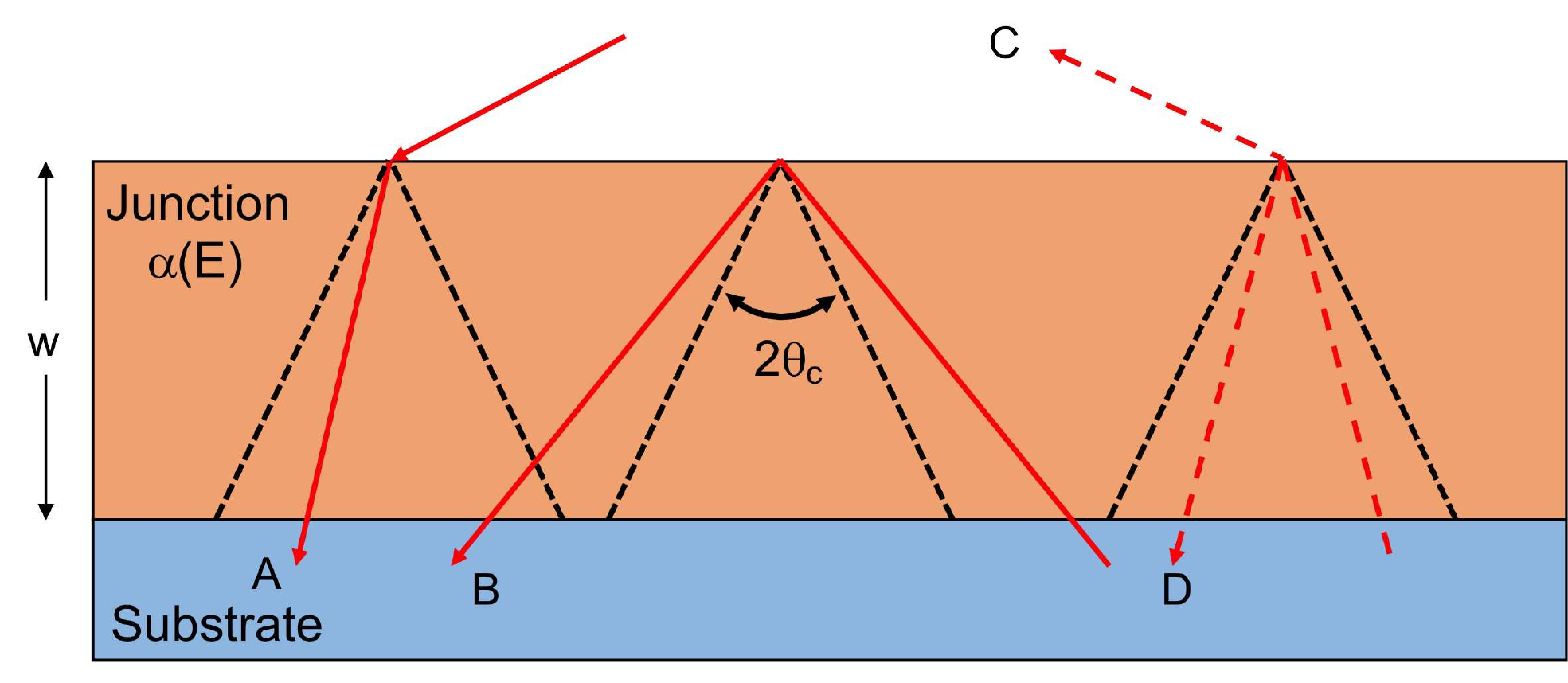}
  	\caption{All the paths that radiation can follow withing the cell, used to calculate the absorptivity$\equiv$emissivity as a function of the angle.}
  	\label{fig:rad_emission}
\end{figure}

The different paths of the absorbed light are depicted in Fig. \ref{fig:rad_emission}. Path A represents light that reaches the front surface within the escape cone ($\theta < \theta_c$) and that crosses the structure. Path B is the light that reaches the back surface outside the escape cone of the front surface ($\theta > \theta_c$), being totally internally reflected and crossing the structure twice. Light reaching the back surface within the escape cone ($\theta < \theta_c$) can either escape through the front (path C) or be reflected (path D). With these considerations, the contribution to the surface integral of the four terms will be~\cite{Nelson:1997fb}:

\begin{strip}
\begin{equation}\label{eq:rad_components}
\begin{split}
\text{A - } & 2\pi S_{front} \int_{\cos \theta_c}^{1} \left[ 1-r(E,\theta) \right] \left( 1 - e^{-\alpha w / \cos \theta}  \right) \cos \theta d(\cos \theta) \\
\text{B - } & 2\pi S_{back} \int_{0}^{ \cos \theta_c } \left( 1 - e^{-2\alpha w / \cos \theta}  \right) \cos \theta d(\cos \theta) \\
\text{C - } & 2\pi S_{back} \int_{\cos \theta_c}^{1} \left[1-r(E, \theta) \right] \left( 1 - e^{-\alpha w / \cos \theta}  \right) \cos \theta d(\cos \theta) \\
\text{D - } & 2\pi S_{back} \int_{\cos \theta_c}^{1} r(E, \theta) \left( 1 - e^{-2\alpha w / \cos \theta}  \right) \cos \theta d(\cos \theta)
\end{split}
\end{equation}
\end{strip}

These equations can be written in a more compact form by noting that $r=1$ for $\theta > \theta_c$. In this situation, B, C and D can be combined and the integral extended from 0 to 1, resulting simply in:

\begin{equation}\label{eq:rad_components_back}
\begin{split}
A_{back}(E) &= 2\pi \int_{0}^{1} \left[1+r(E, \theta) e^{-\alpha w / \cos \theta} \right] \\
&\times \left( 1 - e^{-\alpha w / \cos \theta}  \right) \cos \theta d(\cos \theta)
\end{split}
\end{equation}

The factor $S_{back}$ representing the area of the back of the cell has been omitted here as we are interested in the current density, and therefore independent of the area. Likewise, $A_{front}(E)$ will be given simply by the component A in Eq. \ref{eq:rad_components}:
 
\begin{equation}\label{eq:rad_components_front}
\begin{split}
A_{front}(E) &= 2\pi \int_{\cos \theta_c}^{1} \left[ 1-r(E,\theta) \right] \\
&\times \left( 1 - e^{-\alpha w / \cos \theta}  \right) \cos \theta d(\cos \theta)
\end{split}
\end{equation}

$J_{rad}(V, T_{cell})$ will represent the radiative recombination of the cell at a bias $V$ and temperature $T_{cell}$ while $J_{rad}(0, T_{a})$ will be the carrier generation due to thermal radiation from the ambient at a temperature $T_a$. Typically, $T_{cell} = T_a$. 

Carrier generation in the solar cell due to the absorption of the solar irradiance $H(E)$ can be written simply as:

\begin{equation}
J_{sc} = q \int_{0}^{\infty } [1-R(E)] A_n(E) H(E) dE
\end{equation}

\noindent where $R = r(E, 0)$ is the normal incidence reflection and $A_n$ is the normal incidence absorptivity of the cell, given by $A(E) = 1- \exp (-\alpha(E) w)$ where $w$ is the thickness of the junction and $\alpha$ is the absorption coefficient. 

Combining all these equations, the total current as a function of the bias calculated with the DB model will be given by:

\begin{equation} \label{eq:DB}
J = J_{sc} + J_{rad}(0, T_{a}) - J_{rad}(V, T_{cell})
\end{equation}

If $T_a = T_{cell}$ and $E >> k_bT$, Eq. \ref{eq:DB} simplifies, resulting in: 

\begin{equation}
J = J_{sc} - J_{01} \left( e^{\frac{qV}{k_bT_{cell}} } -1 \right)
\end{equation}

\noindent with  $J_{01}$ the reverse saturation current, given by:

\begin{equation}\label{eq:J01_DB}
J_{01} = q \frac{2 n^2}{h^3c^2}  \int_{0}^{\infty } E^2e^{ -\frac{E}{k_bT_{cell}}} \left[ A_{front}(E)  + A_{back}(E)  \right] dE
\end{equation}

While in these equations the term $\exp(-\alpha w)$ is used, it should be noted that none of the two parameters $alpha$ and $w$ are needed as the product is calculated internally by Solcore from the normal incidence absorptivity $A_n(E)$, which is the value given as input:

\begin{equation}\label{eq:absorptivity}
e^{-\alpha w} = 1-A_n(E)
\end{equation}

\subsection{2-diode model (2D)}

This is the simplest method for simulating the behaviour of a solar cell, using electrical components to model the different transport and recombination mechanisms of the device. The 2D model is widely applied when modelling solar cells at the most engineering end of the topic, when a detailed knowledge of the solar cell structure (layers, absorption coefficient, etc.) are not known or sought. It is often used to fit experimental IV curves and find approximate, general information on the solar cell quality without entering on the fundamental processes. It can provide valuable information to engineers, when designing solar modules for example, or for diagnostic purposes The complete form of the equation is:

\begin{equation} \label{eq:2diode}
\begin{split}
J = J_{sc} & - J_{01} \left( e^ \frac{q(V-R_sJ)}{n_1 k_b T} - 1 \right) \\
& - J_{02} \left( e^ \frac{q(V-R_sJ)}{n_2 k_b T} - 1 \right) \\
& - \frac{V-R_sJ}{R_{sh}}
\end{split}
\end{equation}

Generally, the photocurrent is modelled as a current source ($J_{sc}$), with radiative and non-radiative recombination modelled as two diodes with reverse saturation currents $J_{01}$ and $J_{02}$, and ideality factors $n_1\approx 1$ and $n_2\approx 2$, respectively. The shunt resistance $R_{sh}$ accounts for alternative current paths between the contacts of the solar cell, being infinite in the ideal case, and the series resistance $R_s$ accounts for the other transport losses. The values of the saturation currents and ideality factors can, ultimately, be calculated from the material properties and device structure, as is done in the depletion approximation model (Section~\ref{sec:DAmodel}), but the 2D model allows them to be provided directly as input, obtained from a fit to experimental data, for example. They can also be calculated internally, using the DB solver to obtain $J_{01}$ and $J_{sc}$, and then using a radiative efficiency coefficient to obtain $J_{02}$. The radiative efficiency $\eta$ is defined as the fraction of radiative current $J_{rad}$ at a given reference total current $J_{ref}$:

\begin{equation}\label{eq:eta_rad}
\eta = \frac{J_{rad}}{J_{ref}} = \frac{J_{01}}{J_{ref}} \left( e ^{\frac{qV_{ref}}{n_1k_bT}} - 1 \right)
\end{equation}

The reference voltage $V_{ref}$ can be written as a function of $\eta$ and $J_{ref}$ as:

\begin{equation}\label{eq:vref}
V_{ref} = \frac{n_1k_bT}{q} \log \left( \frac{\eta J_{ref}}{J_{01}} + 1 \right) 
\end{equation}

On the other hand, the radiative coefficient can also be written as:

\begin{equation}\label{eq:eta_nrad}
\eta = \frac{J_{ref} - J_{nrad} - V_{ref}/R_{sh}}{J_{ref}}
\end{equation}

Combining Eq. \ref{eq:vref} and \ref{eq:eta_nrad} and using the expression for the diode with ideality factor $n_2$, $J_{02}$ can be written as:

\begin{equation}\label{eq:J02}
J_{02} = \frac{(1-\eta) J_{ref} - V_{ref} / R_{sh}}{e^ {\frac{qV_{ref}}{n_2k_bT} } - 1 }
\end{equation}

In the common situation of very large shunt resistance and $V_{ref} >> k_bT/q$, this equation further simplifies to:

\begin{equation}\label{eq:J02_simple}
J_{02} = (1-\eta) J_{ref} \left( \frac{J_{01} }{ J_{ref} \eta } \right)^{n_1/n_2}
\end{equation}

This process can, of course, be reversed to use knowledge of $J_{01}$ and $J_{02}$ at a given reference current to calculate the radiative efficiency of a solar cell, which is useful to compare different materials, technologies or processing methods. This was done by Chan et al. using $J_{ref} = 30$ mA/cm$^2$, obtaining $\eta$ values of 20\% for InGaP, 22\% for GaAs, and 27\% for InGaAs devices~\cite{Chan:2012ej}. It should be pointed out that this method is only valid  under the assumption that $J_{01}$ corresponds only to radiative recombination and $J_{02}$ only to non-radiative recombination, which is generally true for QW solar cells and some III-V solar cells, like those made of GaAs or InGaP, but not for Si or Ge, for example. Other definitions of the radiative efficiency are based on the external quantum efficiency, the I$_{sc}$ and V$_{oc}$ of the cell, as described by Green~\cite{Green:2011ea}.

Despite the simplicity of the 2-diode model, it is very useful to guide the design of new solar cells and explore the performance of new materials, such as dilute bismuth alloys~\cite{Thomas:2015ke}, or to asses the performance of large arrays of solar cells, as will be shown in Section~\ref{sec:large scale}~\cite{ekins:04b}.

\subsection{Poisson-drift-diffusion (PDD)}\label{sec:PDD}

This method solves the Poisson equation for the electrostatic potential coupled with the transport equations for electrons and holes and suitable boundary conditions in the steady state. It is the standard method for calculating the electrical properties of most semiconductor devices, including solar cells, transistors or light emitting diodes. It is also the only method included in most software packages for simulating semiconductor devices, such as PC-1D, Nextnano, SCAPS and AFORS-HET. 

Figure \ref{fig:PDD}a shows the flow chart of Solcore's PDD solver, which currently only solves the time-independent PDD equations (steady state). Any simulation starts by calculating the band structure under equilibrium conditions (no illumination or bias). If the simulation includes illumination, the photogeneration as a function of the position in the structure is calculated externally to the PDD solver using any of the models described in Section \ref{sec:Optical solvers}. To aid convergence, the solution at short circuit conditions is calculated by increasing the light intensity from zero to the nominal value in small steps. Similarly, the solution at any bias is obtained by solving the problem first at zero bias and then increasing it in small steps, using the previous solution as the initial condition for the next one. Re-meshing is performed several times during the simulation of the current-voltage characteristics (see section \ref{sec:meshing}). 

To calculate the internal quantum efficiency (IQE), a small differential increase is included in the photogeneration profile as a function of wavelength. The IQE is then calculated as the ratio between the resulting increase in the photocurrent and the increase in the photogeneration at that wavelength. This procedure is comparable to the actual experimental measurement of the quantum efficiency. 

\subsubsection{Solver assumptions and formulation}

The Poisson's and drift-diffusion equations relate the electrostatic potential created by the free and fixed charges with the carrier densities and their variation across the structure due to generation, recombination and externally applied bias. The reader is referred to any semiconductor textbook for a detailed description of their derivation (see for example~\cite{Sze:100213} or ~\cite{Nelson:2003vl}). The solver uses the Boltzmann approximation for the carrier distribution with the following assumptions:

\begin{itemize}
\item All carrier populations are in quasi-thermal equilibrium.
\item The mobility of carriers is independent of the electric field. 
\item Temperature is uniform.
\item There are no magnetic fields.
\end{itemize}

As a consequence of the field-independent mobility, Solcore's PDD solver will be valid only in situations where electric field is not very strong. Poisson's equation relates the electrostatic potential $\phi$ and the electrical charges in the structure. In one dimension, it can be written as:

\begin{equation} \label{eq:poisson}
\frac{d}{dx} \left( \epsilon \frac{d \phi }{d x} \right) + q \left( p - n + N_D - N_A \right) = 0
\end{equation}

\noindent where $N_A$, $N_D$, $n$ and $p$ are the density of ionized acceptors, donors, the density of free electrons and holes, respectively, and $\epsilon$ the dielectric constant. The current density equations account for the movement of carriers due to the electric field (drift component) and the carrier concentration gradient (diffusion component):

\begin{equation} \label{eq:drift_diffusion_n}
J_n = q \mu_n \left( n F + \frac{k_b T}{q} \frac{dn}{dx} \right)
\end{equation}

\begin{equation} \label{eq:drift_diffusion_p}
J_p = q \mu_p \left( p F + \frac{k_b T}{q} \frac{dp}{dx} \right)
\end{equation}

\noindent where $\mu$ is the carrier mobility and $F= - d \phi / dx$ the electric field. Finally, the continuity equations ensure particle conservation, balancing the carriers that enter and leave any point of the structure. Under steady state, this means that the variation of the current must be equal to the generation $G$ and recombination $R$ processes, which are equal for electrons and holes since they are created and annihilated in pairs:

\begin{equation} \label{eq:continuity_n}
\frac{d J_n}{dx} + q G - q R = 0
\end{equation}

\begin{equation} \label{eq:continuity_p}
-\frac{d J_p}{dx} + q G - q R = 0
\end{equation}

Combining Eq. \ref{eq:drift_diffusion_n}, \ref{eq:drift_diffusion_p}, \ref{eq:continuity_n} and \ref{eq:continuity_p} gives:

\begin{equation} \label{eq:DD_n}
q \mu_n \left( \frac{k_b T}{q} \frac{d^2n}{dx^2} + F \frac{dn}{dx} + n\frac{dF}{dx}  \right)+ q G - q R = 0
\end{equation}

\begin{equation} \label{eq:DD_p}
q \mu_p \left( \frac{k_b T}{q} \frac{d^2p}{dx^2} - F \frac{dp}{dx} - p\frac{dF}{dx}   \right)+ q G - q R = 0
\end{equation}

Poisson's equation (Eq. \ref{eq:poisson}) and the continuity equations (Eq. \ref{eq:DD_n} and Eq. \ref{eq:DD_p}), together with the definitions for $n$, $p$ and $R$, represent the complete system that needs to be solved in order to obtain the performance of the solar cell. The PDD solver included in Solcore uses the same discretization scheme used by PC-1D~\cite{Basore:wP5Wjg, Basore:62P3ZbQB} taking as independent variables the electrostatic potential $\phi$ and the quasi-Fermi potentials for electrons and holes, $\phi_{n}$ and $\phi_{p}$, respectively. These three variables are continuous across the whole structure and have comparable magnitudes in the voltage range. Details of the discretization process are included in ~\cite{Basore:62P3ZbQB, Basore:wP5Wjg}, but they are based on minimising the total electrostatic energy in the case of the Poisson's equation, and in the Scharfetter-Gummel discretization scheme for the drift-diffusion equations ~\cite{Farrell:1991vr}.

The bulk recombination models included in Solcore are Shockley-Read-Hall recombination, radiative recombination and Auger recombination, as well as a surface recombination velocity model for the recombination at the contacts. The three recombination models are given by the following equations, respectively:

\begin{equation} \label{eq:SRH}
R_{SRH} = \frac{pn-n_i^2}{\tau_n (p+n_i) + \tau_p (n+n_i)}
\end{equation}

\begin{equation} \label{eq:RAD}
R_{RAD} = B(pn-n_i^2)
\end{equation}

\begin{equation} \label{eq:AUG}
R_{AUG} = ( C_n n + C_p p ) (pn-n_i^2)
\end{equation}

\noindent with $\tau_n$ and $\tau_p$ the non-radiative lifetimes of electrons and holes, $B$ the radiative recombination coefficient and $C_n$ and $C_p$ the Auger recombination coefficient for electrons and holes. By default, the radiative recombination coefficient is calculated internally by Solcore based on the absorption coefficient, as described by Nelson~\cite{Nelson:2003vl}, and given by:

\begin{equation} \label{eq:Br}
B = \frac{1}{n_i^2}\frac{2 \pi}{h^3 c^2}  \int_{0}^{\infty } n_s(E)^2 \alpha(E) E^2e^{ -\frac{E}{k_bT}} dE 
\end{equation}

\noindent where $n_i$ is the intrinsic carrier concentration, $n_s$ the refractive index and $\alpha$ the absorption coefficient. 

At the moment, Solcore's PDD solver cannot include interface charges or bandgap narrowing due to heavy doping and only implements Ohmic contacts. Additionally, it only includes local carrier recombination processes and therefore cannot deal with tunnel transport, which relates remote nodes. 

This is the only part of Solcore implemented in Fortran using quadruple precision variables in order to increase the numerical accuracy and improve convergence. 

\subsubsection{QWs in the PDD solver}

\begin{figure}
  	\centering
  	\includegraphics[width=\columnwidth]{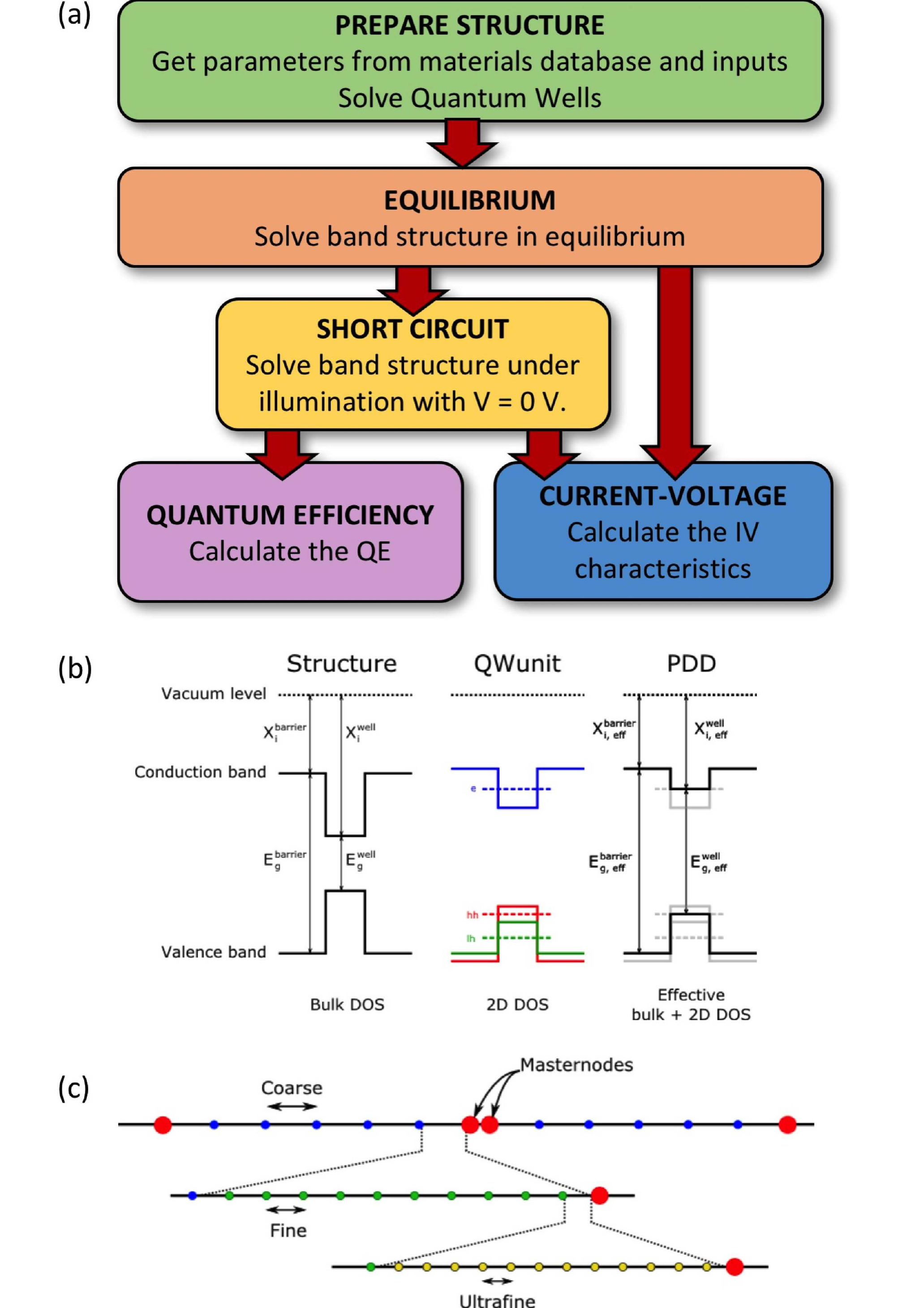}
  	\caption{(a) Work flow of Solcore's PDD solver. (b) Process of obtaining the effective band structure of QWs to use in the PDD solver. From left to right: simple sequence of layers; band profile and energy levels after considering the strain and quantum confinement; effective band structure. (c) Description of the inhomogeneous mesh scheme used in Solcore.}
  	\label{fig:PDD}
\end{figure}

Quantum wells have been developed in the context of solar cells mainly to tailor the absorption edge of the sub-cells in multi-junction devices to their optimum values~\cite{Thomas:2014ta}. Typically, achieving the proper performance requires a delicate trade-off between carrier collection and light absorption ~\cite{AlonsoAlvarez:2014gg, AlonsoAlvarez:2016ki} . Solcore includes a simplified QW structure in the PDD solver in order to calculate the performance of solar cells containing them. Contrary to other programs like Nextnano, Solcore does not solve the Schr\"odinger equation and the PDD equations self-consistently: first, the energy levels of the quantum wells are solved using a flat-band condition, considering also the strain in the materials, and then an effective band structure is used to solve the transport equations in a bulk-like fashion. This is illustrated in Figure \ref{fig:PDD}b.

From the perspective of the PDD solver, the actual bandgap and electron affinity of each layer in a quantum well depend on the energy levels, i.e. the minimum energy for electrons is not the band edge of the conduction band, but the ground confined level. The same applies to holes, with the actual band edge being the maximum between the ground states of light holes and heavy holes. The resulting band profiles used in the PDD solver are shown in the right picture of Figure \ref{fig:PDD}b.

To use QWs in the PDD solver, we create an effective electron affinity and bandgaps for all layers in the QW. For the barriers, the electron affinity and band gap are the same as they are in bulk, modified by the strain, if necessary. For interlayers, if present, it depends on what is higher, the band edges of the interlayer or the confined carrier levels.

The density of states and the absorption profile need to be modified in a similar way. For the density of states: 

\begin{itemize}
\item \textbf{Barriers} have the bulk density of states and absorption profile. 
\item \textbf{Interlayers} only have the bulk density of states above the barrier and the bulk absorption from the barrier energy and zero below that.
\item \textbf{Wells} have all the density of states associated with the confined states and the bulk density of states above the barrier, while they have the absorption of the confined levels below the barrier energy and of the bulk above it.
\end{itemize}

These simplifications are similar to those in Nelson et al. ~\cite{Nelson:2003vl} and in Cabrera et al. ~\cite{Cabrera:2013gi} and allow us to keep the bulk-like form of the carrier densities in the drift diffusion equations under the Boltzmann approximation. A more rigorous treatment will be necessary in the presence of tunnel transport across a supperlattice, tunnel escape from the QWs to the barriers - possible in the presence of high electric fields - and in the case of very deep QWs, when carrier escape from the less confined levels might be possible but not from the deeper ones. In these situations, a set of rate equations linking the different levels, as well as a self-consistent solution of the transport and Schr\"odinger equations would be required, besides using more advanced methods such as a non-equilibrium Green's functions (NEGF) formalism~\cite{Aeberhard:2018it}. 

\subsubsection{Mesh creation and dynamic meshing}\label{sec:meshing}

The PDD solver discretizes the device into a finite number of mesh points at which to calculate the band structure, carrier densities, the generation and recombination. The mesh can be static and homogeneous, but the default configuration and the one that results in the least number of mesh points, most accurate result and best convergence uses an inhomogeneous mesh with dynamic re-meshing. 

There are two types of nodes in the mesh: masternodes and normal nodes. There are two masternodes at the ends of the device and two more at each side of an interface separated by 0.1 nm. These nodes are static and not affected by re-meshing. The rest of the nodes are automatically distributed depending on the distance to the closest masternode, as illustrated in Figure \ref{fig:PDD}c. During re-meshing, nodes can be added or removed according to the following rules:

An element will be divided into smaller elements by adding new nodes if any of the following statements is true:

\begin{itemize}
\item The variation of the potentials or the carrier densities across the element is large.
\item The element is too close to the masternodes limiting the layer.
\item The element is too big for the region.
\end{itemize}

A node will be removed if it fulfils all the following conditions:

\begin{itemize}
\item It is not a masternode.
\item The variation of the potentials or the carrier densities with respect the previous and next nodes is small.
\item It is not too close to the masternodes limiting the layer.
\item Removing it does not create an element too big for the region.
\end{itemize}

After the initial meshing and every time there is a re-meshing, the position of the nodes (except that of the masternodes) is smoothed to avoid having adjacent elements too different in size. This re-meshing process is controlled by a \textit{growth} parameter, which can be adjusted by the user. 

Using the inhomogeneous mesh in addition to the dynamic re-meshing ensures that those regions where material properties change abruptly are modelled with more detail, aiding the convergence. It also allows the modelling of devices which have layers with very different thickness, such as QWs a few nanometers thick and bulk absorbers of several microns, without increasing the number of nodes significantly. 

\subsection{Depletion approximation}\label{sec:DAmodel}

The depletion approximation provides an analytical - or semi-analytical - solution to the Poisson-drift-diffusion equations described in the previous section applied to simple PN homojunction solar cells. Historically, it has been used extensively to model solar cells and it is still valid, to a large extent, for traditional PN junctions. More importantly, it requires less input parameters than the PDD solver and these can be easily related to macroscopic measurable quantities, like mobility or diffusion lengths. The DA model is based on the assumption that around the junction between the P and N regions, there are no free carriers and therefore all the electric field is due to the fixed, ionized dopants. This ``depletion'' of free carriers reaches a certain depth towards the N and P sides; beyond this region, free and fixed carriers of opposite charges balance and the regions are neutral. Under these conditions, Poisson's equation decouples from the drift and diffusion equations and it can be solved analytically for each region. For example, for a PN junction with the interface between the two regions at $z=0$, the solution to Eq. \ref{eq:poisson} will be:

\begin{equation}
\phi(z) =
\left\{
	\begin{array}{ll}
		0  & \mbox{if } z < -w_p \\
		\frac{qN_a}{2\epsilon_s}(z+w_p)^2  & \mbox{if } -w_p < z < 0 \\
		-\frac{qN_d}{2\epsilon_s}(z-w_n)^2 + V_{bi} & \mbox{if } 0 < z < w_n  \\
		V_{bi} & \mbox{if } w_n < z
	\end{array}
\right.
\end{equation}

\noindent where $w_n$ and $w_p$ are the extensions of the depletion region towards the N and P sides, respectively, and can be found by the requirement that the electric field $F$ and the potential $\phi$ need to be continuous at $z=0$. $V_{bi}$ is the built-in voltage, which can be expressed in terms of the doping concentration on each side, $N_d$ and $N_a$, and the intrinsic carrier concentration in the material, $n_i^2$:

\begin{equation}
V_{bi} = \frac{k_bT}{q} \ln \left(\frac{N_dN_a}{n_i^2} \right)
\end{equation}

Another consequence of the depletion approximation is that the quasi-Fermi level energies are constant throughout the corresponding neutral regions and also constant in the depletion region, where their separation is equal to the external bias $qV$. Based on these assumptions, Eq. \ref{eq:drift_diffusion_n} and \ref{eq:drift_diffusion_p} simplify and an analytical expression can be found for the dependence of the recombination and generation currents on the applied voltage. A full derivation of these expressions is included in Nelson~\cite{Nelson:2003vl}.

Solcore's implementation of the depletion approximation includes two modifications to the basic equations. The first one is allowing for an intrinsic region to be included between the P and N regions to form a PIN junction. For low injection conditions (low illumination or low bias) this situation can be treated as described before, simply considering that the depletion region is now widened by the thickness of the intrinsic region. Corrently, no low doping level is allowed for this region.

The second modification is related to the generation profile, which in the equations provided by Nelson is given by the BL law (Eq. \ref{eq:BL}) which has an explicit dependence on $z$ and results in analytic expressions for the current densities. In Solcore, we integrate the expressions for the drift-diffusion equations under the depletion approximation numerically to allow for an arbitrary generation profile calculated with any of the methods described in Section \ref{sec:Optical solvers}. It should be noted that although the equations are integrated numerically this will not be a self-consistent solution of the Poisson-drift-diffusion equations, as is achieved by the PDD solver (Section \ref{sec:PDD}).

\section{Multi-junction solar cells}\label{sec:MJ}

A complete photovoltaic solar cell can include one or more junctions, metal contacts, optical layers (including anti-reflective coatings and nano-photonic structures) and tunnel junctions. The junctions, in turn, might range from simple PN homojunctions to complex heterojunctions, including multi-quantum well structures. The solvers described in Section \ref{sec:electrical solvers} only calculate the properties of single junction devices. To combine them into a multi-junction device, it is necessary to consider that the individual junctions are electrically connected in series and the potential coupling of light emitted by the wider bandgap junctions into those with smaller bandgap. The Suplementary Information includes a full step-by-step example of the modelling of a dual junction solar cell with QWs, anti-reflecting coating and a tunnel junction, calculating the external quantum efficiency, the IV characteristics under illumination and and the performance of the solar cell as a function of light concentration. 

\subsection{No radiative coupling}

We first consider the case of no radiative coupling between junctions. This is a good approximation for materials which do not radiate efficiently or radiative materials working at low concentration, when the fraction of radiative recombination compared to non-radiative recombination is low. In this case, the IV curve of each junction can be calculated independently of each other and the current flowing through the MJ structure is limited by the junction with the lowest current at any given voltage. Series resistances defined for each junction are now added together and included as a single term. The operating voltage of each of the junctions is finally back-calculated and added together to get the voltage of the MJ device. 

The pseudocode for this solver is:

\begin{enumerate}
\item Calculate the $I_j(V)$ of each junction $j$ in the structure.
\item Find the current flowing through the MJ device as $I_{MJ}(V) = I_j(V)$, if $|I_j(V)| = \min ([|I_1(V)|...|I_N(V)|])$.
\item Calculate the voltage of each junction by interpolating its IV curve at the new current values, $V_j(I_{MJ})$, and the voltage dropped due to the series resistances, $V_{Rs} = R_s I_{MJ}$.
\item Calculate the total voltage at a given current as $V_{MJ} = V_{Rs} + \sum_j V_j$. 
\item Interpolate the $I_{MJ}(V_{MJ})$ and the $I_{MJ}(V_j)$ to the desired output voltage values. 
\end{enumerate}

\begin{figure}
  	\centering
  	\includegraphics[width=\columnwidth]{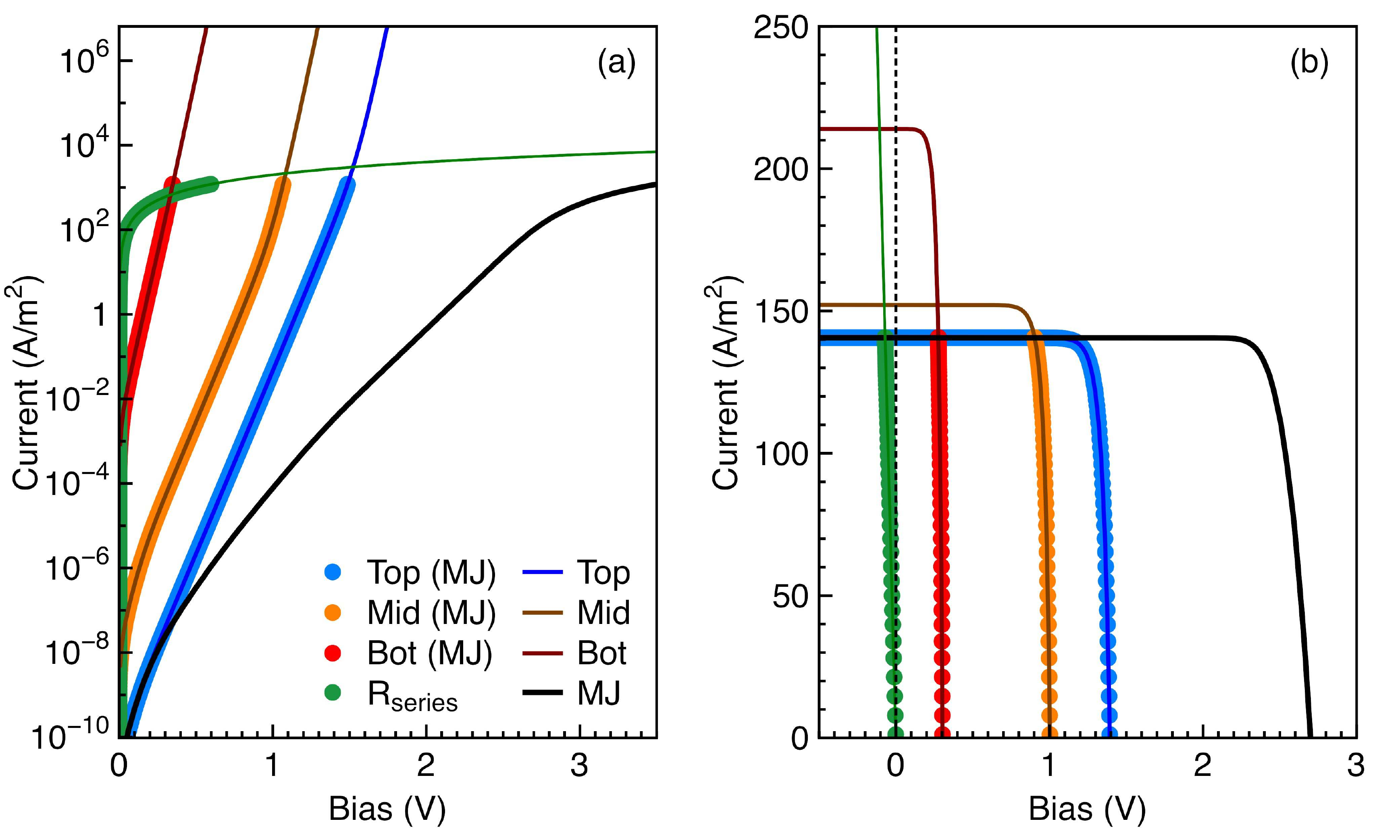}
  	\caption{(a) Dark IV curve of a MJ solar cell, including the IV of the individual junctions separately (continuous lines) and the junctions as part of the MJ structure. (b) Light IV curve of the same MJ solar cell.}
  	\label{fig:MJ_iv}
\end{figure}

Fig. \ref{fig:MJ_iv} shows the simulated IV curve of a 3J solar cell made of Ge/InGaAs/GaAsP. The electrical properties of the three junctions were calculated using the depletion approximation solver. In the dark (Fig. \ref{fig:MJ_iv}a) the voltages of each of the junctions at a given current add together, resulting in the total voltage of the MJ structure. The $R_s$ contribution to the voltage goes in the same direction as those of the junctions. Under illumination (Fig. \ref{fig:MJ_iv}b) the junction producing the lower current (the top junction in this case) limits the overall current of the MJ device. At zero bias, or even at some negative bias, the non-limiting junctions are positively biased, recombining all the photocurrent that cannot be extracted because of the limiting top cell. The contribution of the $R_s$ to the voltage of the MJ device is negative, resulting in a reduction of the fill factor and the overall efficiency of the solar cell. 

\subsection{With radiative coupling}

Radiative coupling takes place when the light produced by a high bandgap junction due to radiative recombination is absorbed by a lower bandgap junction, contributing to its photocurrent and changing the operating point. It has been identified in numerous highly radiative materials such as quantum well solar cells and III-V MJ solar cells~\cite{Lee:2012ee, Steiner:2013cr, Steiner:2012ce}. It appears as an artefact during the QE measurements of MJ solar cells ~\cite{Steiner:2012ip}, but it is also an effect that can be exploited to increase the performance of MJ devices ~\cite{Thomas:2014ta} and their tolerance to spectral changes, resulting in superior annual energy yield ~\cite{Chan:2014ca}.

The radiative coupling formalism included in Solcore is based on the works by Chan et al. and Nelson et al. ~\cite{Chan:2014ca, Nelson:1997fb}. It is implemented only for the DB junction model and for the 2D model when it is defined in terms of a radiative efficiency and the parameters calculated form the DB model. The current of a junction $j$ including radiative coupling from the junction immediately above it $j-1$ is given by:

\begin{equation}
J_j^{total} = J_j^{nc} + J_{j-1 \rightarrow j}^{coupled}
\end{equation}

This current depends on two factors: the amount of radiation effectively emitted downwards, towards the lower junction, and the fraction of it that is absorbed and converted into electricity. If we ignore the possible reflection of light at the interface between both junctions, this current can be written by using a modified version of Eq. \ref{eq:DB_radiative} that considers only the radiation emitted towards the back:

\begin{equation}
\begin{split}
J_{j-1 \rightarrow j}^{coupled}(V, T) &= q \frac{2 n^2}{h^3c^2}  \int_{0}^{\infty } \frac{E^2}{e^{ \frac{E - qV}{k_bT}} - 1 }  A_{j-1 \rightarrow j}(E) dE
\end{split}
\end{equation}

\noindent with $A_{j-1 \rightarrow j}(E)$ given by:

\begin{equation}\label{eq:rad_coupling}
\begin{split}
A_{j-1 \rightarrow j}(E) &= 2\pi \int_{0}^{1} \left[1+r(E, \theta) e^{-\alpha_{j-1} w_{j-1} / \cos \theta} \right] \\
&\times \left( 1 - e^{-\alpha_{j-1} w_{j-1} / \cos \theta}  \right) \\
&\times \left( 1 - e^{-\alpha_j w_j / \cos \theta}  \right) \cos \theta d(\cos \theta)
\end{split}
\end{equation}

As discussed previously, any information related to total internal reflection will be contained in the $r(E, \theta)$ term, and therefore the integral over $\cos \theta $ can be done from 0 to 1. In the case of thin junctions, some radiation could reach the next junction $j+1$. The coupled current in that case can be easily calculated by modifying Eq. \ref{eq:rad_coupling} to account for the fraction of light absorbed by the junctions between the emitting junction and the junction of interest. In the general case, the current coupled into junction $j$ will be given by:

\begin{equation}
J_{j}^{coupled} = \sum_{i=1}^{j-1} J_{i \rightarrow j}^{coupled}
\end{equation}

Radiative coupling might change the junction that is current limiting the MJ device, so the process to obtain the IV curves in this case proceeds in two steps. First, the IV of the junctions and the total IV are calculated without coupling. The resulting IV curves are then used as the initial conditions for the numerical solver that will calculate the correct voltage of each junction including the radiative coupling. 

\begin{figure}
  	\centering
  	\includegraphics[width=\columnwidth]{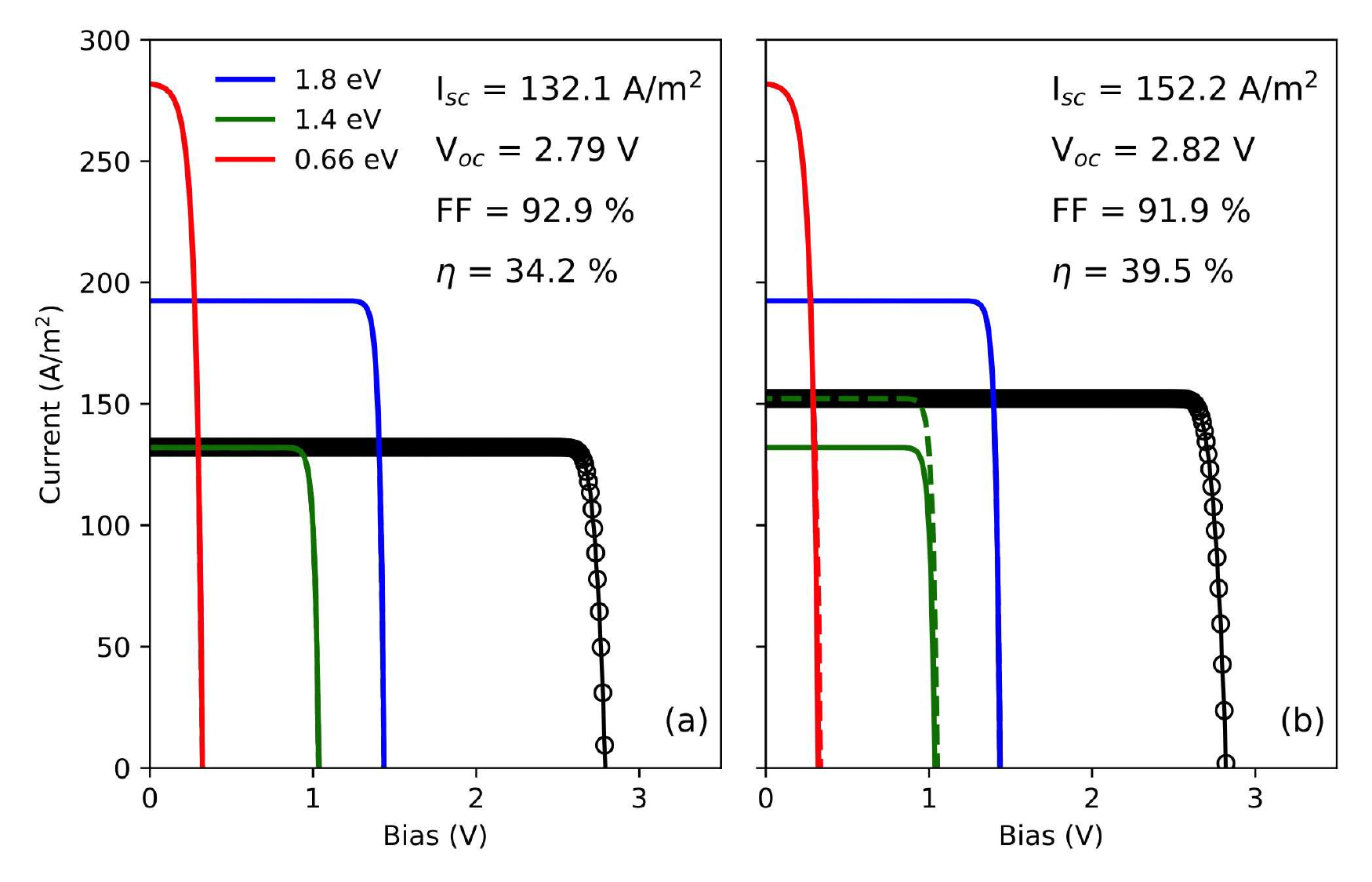}
  	\caption{Light IV curve of a 3J solar cell (a) without and (b) with radiative coupling. Continuous lines represent the individual IV curves of the junctions isolated and the dash lines when they are inside the MJ device, illustrating the effect of the coupling.}
  	\label{fig:rad_coupling}
\end{figure}

Fig. \ref{fig:rad_coupling}  shows the IV curve under the AM1.5G solar spectrum of a three junction solar cell (a) without and (b) with radiative coupling. Without coupling, the middle junction severely limits the current of the MJ solar cell. When coupling is enabled, the middle junction is still the limiting one but part of the excess current of the top junction is transferred to it, increasing its photocurrent by around 20 A/m$^2$. Part of the radiative recombination is also transferred to the bottom cell, increasing slightly its photocurrent. In this case, given that the junction was overproducing current already, such coupling is only visible as an increase in the voltage. Altogether, the radiative coupling results in an enhancement of the V$_{oc}$ of 30 mV and of the efficiency $\eta$ of 5.3\%. This example uses junctions with 100\% radiative efficiency to illustrate the effect, but this phenomenon is always present to some extent, becoming especially important under concentration~\cite{Thomas:2014ta, Chan:2014ca}.

\subsection{Restrictions in the Junction definitions}

Having multiple methods for modelling the junctions gives a lot of freedom and flexibility but it also imposes some restrictions in how and when they can be combined in order to create a MJ solar cell. The following compatibility rules apply:

\begin{itemize}
\item When there is no radiative coupling and we are interested only in the dark IV characteristics, all junction models can be combined with each other. This allows a MJ device where the top junction is defined using the DB model, the middle junction is defined with the 2D model and the bottom junction uses a more accurate PDD model, for example. 
\item The same applies for light IV and quantum efficiency simulations as long as the optical model chosen is the BL law. In this case, any junction defined using the 2D model needs to include an absorptivity value.
\item The TMM and RCWA optical models are supported only by the PDD and DA junction models. 
\item In the presence of radiative coupling, the only junction models that can be used are DB and 2D, as long as the latter includes an absorptivity value. 
\end{itemize}

\subsection{Tunnel junctions}

Solcore includes partial support for tunnel junctions. They represent an optical loss due to parasitic absorption in the layers, but also an electrical loss. At the moment, there are two models for tunnel junctions. The first one is a simple resistive model, where the  tunnel junction is simply modelled as a series resistance. This approximation should be valid in most cases, but will break down if the current is close to or higher than the peak current density of the junction.  

The second model is a parametric model, based on the simple formalism described by Sze~\cite{Sze:100213}. In this model, the total current of the tunnel junction will have three components: the tunnel current $J_T$ accounting for band-to-band transport, the excess current $J_{ex}$ related to transport across states inside the forbidden gap, and the diffusion current $J_D$, which is the usual minority-carrier injection current in PN junctions. The following equations summarise all these components. 

\begin{equation}
J_{TJ} = J_{T} + J_{ex} + J_D
\end{equation}

\begin{equation}
J_{T} = \frac{J_P V}{V_P} \exp{\left( 1 - \frac{V}{V_P} \right) }
\end{equation}

\begin{equation}
J_{ex} =J_V \exp{\left[ C \left( V - V_V \right) \right] }
\end{equation}

\begin{equation}
J_{D} =J_0 \left[  \exp{ \left( \frac{qV}{k_b T}  \right) } - 1  \right]
\end{equation}

\begin{figure}
  	\centering
  	\includegraphics[width=0.9\columnwidth]{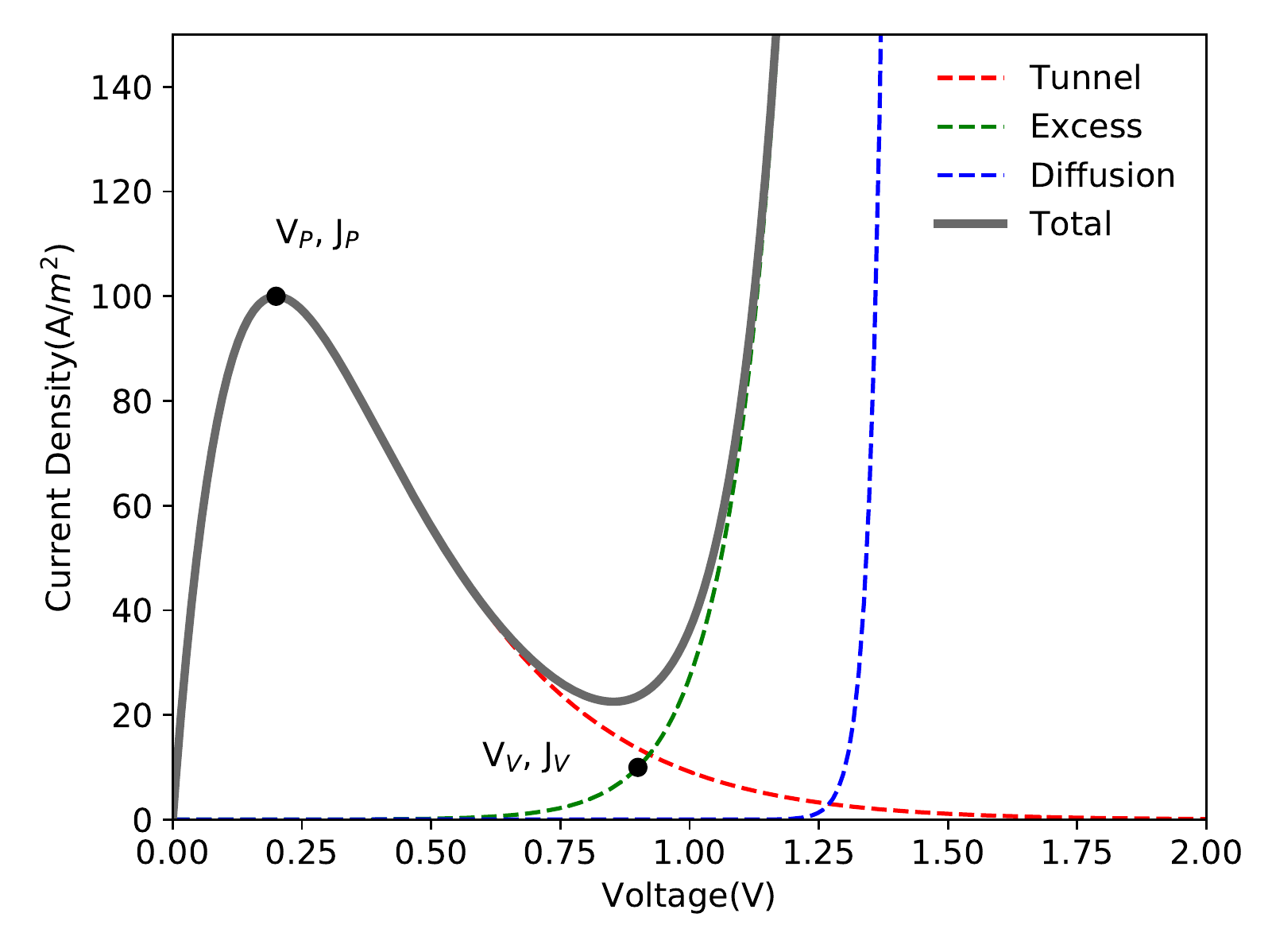}
  	\caption{IV curve of a tunnel junction defined according to the parametric model.}
  	\label{fig:tunnel}
\end{figure}

As illustrated in Fig.~\ref{fig:tunnel}, $J_P$ and $V_P$ are the peak current and voltage, $J_V$ and $V_V$ are the valley current and voltages, $C$ is a prefactor of the exponent and $J_0$ the reverse saturation current. In this simple implementation, these 6 parameters need to be provided as inputs, and can be used as fitting parameters to reproduce experimental data. This allows to correctly account for the break down of the tunnel junction in situations when the current is above the peak current. 

Solcore can also accept external IV data for the tunnel junctions and the implementation of the more rigorous, but still analytic model, described by Louarn \emph{et al.} is currently under way in order to relate the tunnel currents with the actual materials and layer structure used in the tunnel junction definition~\cite{Louarn:2016it}. 

\section{Large circuit solver} \label{sec:large scale}

When the two diode model is used to define the junctions in a MJ solar cell, then larger scale circuits can be constructed. Solcore includes two levels of large scale equivalent circuits: quasi-3D solar cell modelling and solar array modelling. Both solvers are based on the interface between Solcore and SPICE, allowing for a fast calculation of complex structures with many elements. 

\subsection{Quasi-3D solar cell model}

\begin{figure*}
  	\centering
  	\includegraphics[width=0.9\textwidth]{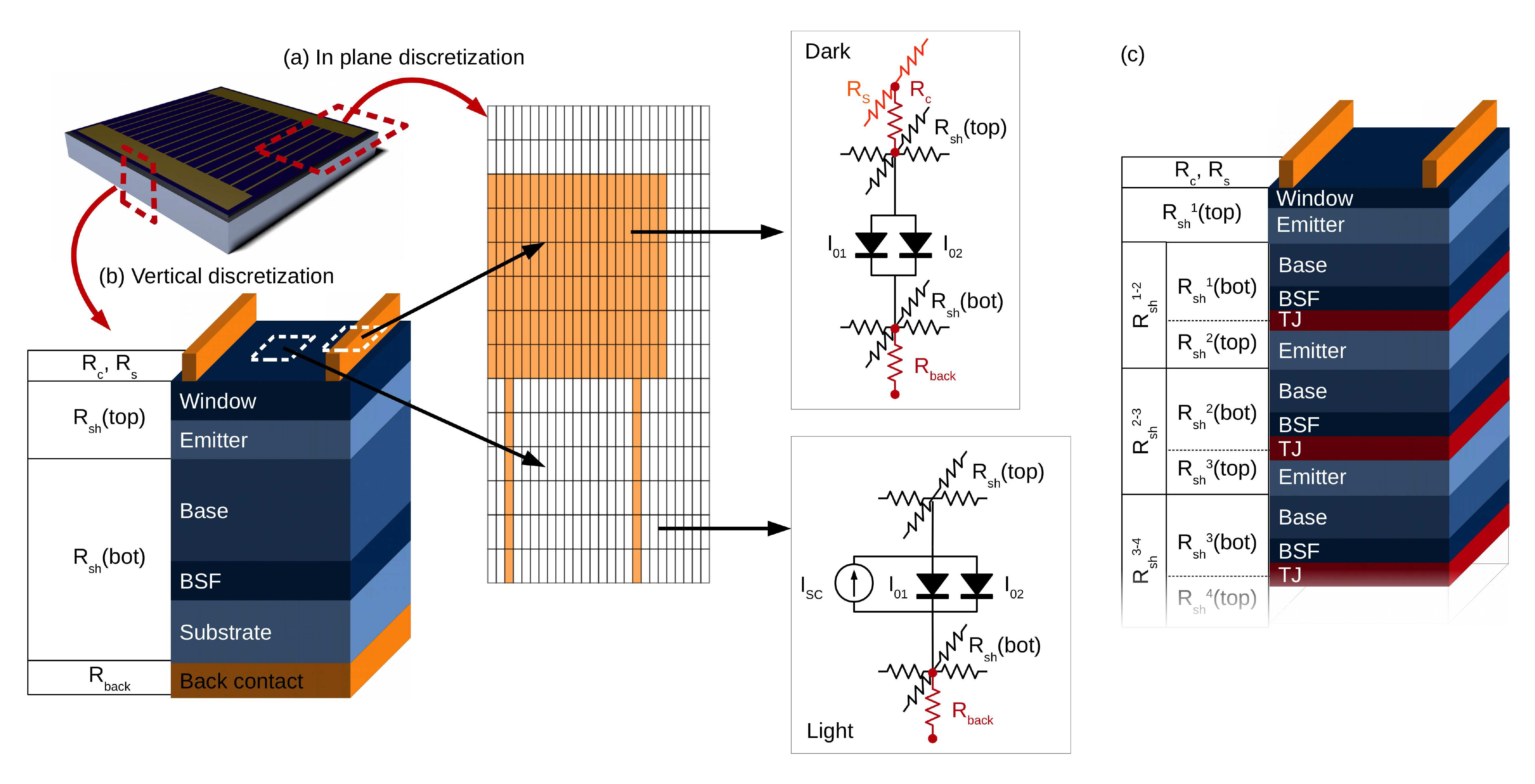}
  	\caption{Schema of the quasi-3D solar cell modelling included in Solcore. The solar cell is discretized (a) in the plane and (b) in the vertical direction. Illuminated and dark regions are then modelled using electrical components that, when combined, form a 3D electrical mesh giving the voltages and currents at any point of the structure. (c) An example of the vertical discretization of a N-junction solar cell.}
  	\label{fig:spice_overview}
\end{figure*}

The quasi-3D solar cell model included in Solcore uses a SPICE-based electrical network to model the flow of injected current through the solar cell, as depicted in Fig. \ref{fig:spice_overview}. The plane of the cell is discretized into many elements, each of them representing a small portion of the cell. Depending on the location of the element - exposed to the sunlight or underneath a metal finger - the IV curve of the cell will be the light IV or the dark IV. Each element is linked to their neighbours with resistors, representing the lateral current flow and dependent on the sheet resistance of the cells. This method can be applied to any number of junctions. 

This type of formalism is widely used to simulate the performance of solar cells when the effect of a spatial variable needs to be incorporated in the model. This variable can be the design of the front metal grid, in order to minimise the effect of series resistances ~\cite{Steiner:2010id}; the inhomogeneous illumination profile in concentrator devices; the impact of such inhomogeneity on the transport through the tunnel junctions ~\cite{Nishioka:2004er, Steiner:2011jy}; or the distribution of defects or inhomogeneities ~\cite{Jurgens:2006gg, Paire:2011co}. Recently, this formalism was used to model the photoluminescence and the electroluminescence based IV curves of MJ devices, accounting for the limited lateral carrier transport ~\cite{AlonsoAlvarez:2016ea}.

Specifically for the modelling and optimization of the front grid of solar cells in order to minimise shading losses and series resistance, there are two packages already available: PVMOS, developed by B. E. Pieters in C and released as open source~\cite{Pieters:2014fg, PVMOS}, and Griddler, developed by J. Wong using Matlab and available at PV Lighthouse~\cite{Wong:2013, PVLighthouse}. 

\subsubsection{In-plane discretization}

\begin{figure}
  	\centering
  	\includegraphics[width=0.9\columnwidth]{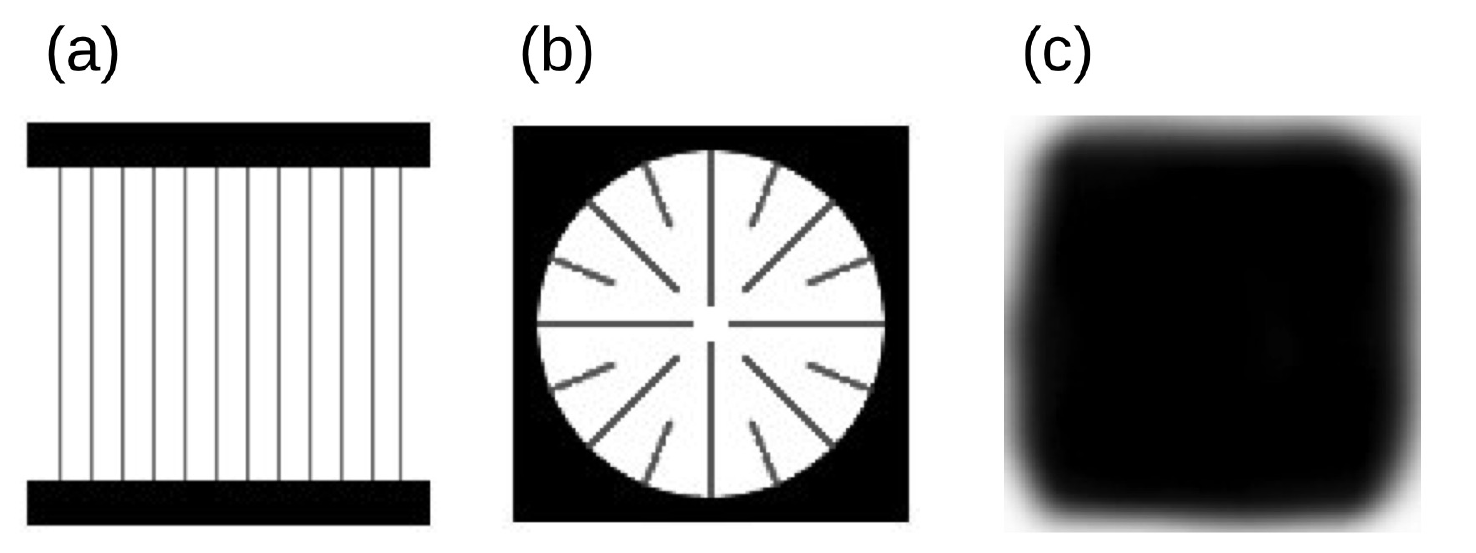}
  	\caption{(a) and (b) two examples of metal masks and (c) an illumination mask. The thin metal fingers in (a) and (b) are grey, indicating that there is metal in those pixels but that their bias is not set to be equal to the external bias.}
  	\label{fig:spice_masks}
\end{figure}

As shown in Fig. \ref{fig:spice_overview}, there are two regions in the plane: the metal and the aperture. These two are provided to Solcore as grey scale images that will work as masks. The resolution of the images, in pixels, will define the in-plane discretization. By default, the aspect ratio of the pixels in the image will be 1:1, but this can be set to a different value in order to reduce the number of elements and improve speed. For example, the in-plane discretization of Fig. \ref{fig:spice_overview}a has an aspect ratio $A_r=L_y/L_x = 4$, with $L_x$ and $L_y$ the pixel size in each direction. The values of the pixels in the metal mask are 0 where there is no metal (the aperture area), 255 where there is metal and the external electrical contacts (the boundaries with fixed, externally set voltage values) and any other value in between to represent regions with metal but not fixed voltage. The pixels of the illumination mask - which become the aperture mask after removing the areas shadowed by the metal - can have any value between 0 and 255. These values divided by 255 will indicate the intensity of the sunlight at that pixel relative to the maximum intensity. Fig. \ref{fig:spice_masks} illustrates two examples of metal masks (a and b) and an illumination mask (c) with 120$\times$120 pixels. As it can be seen, while rectangular metal fingers are well reproduced, diagonal fingers are less accurate and could require a finer discretization. The illumination mask is mostly homogeneous except around the edges and in the corners, where intensity is much lower. This pattern could be produced, for example, by the secondary optics of a concentration system.  

The minimum total number of nodes where SPICE will need to calculate the voltages will be N$\times$M$\times$2$\times$Q, with N and M the number of pixels in both in-plane directions and Q the number of junctions, which require 2 nodes each. To this, the front and back metal contacts could add a maximum of 2M$\times$M nodes. Exploiting symmetries of the problem as well as choosing an appropriate pixel aspect ratio will significantly reduce the number of nodes and therefore the time required for the computation of the problem. 

\subsubsection{Vertical discretization}

First, the solar cell is solved as described in Section \ref{sec:electrical solvers} and \ref{sec:MJ} in order to obtain the parameters for the 2-diode model at the given illumination conditions. These parameters are then used to replicate the 2-diode model in SPICE. The $I_{SC}$ is scaled in each pixel by the intensity of the illumination given by the illumination mask. Sheet resistances above and below each junction, $R_{sh}(top)$ and $R_{sh}(bot)$, account for the lateral transport. Beneath the metal, there is no current source, as the region is in the dark, and there are extra resistances accounting for the contact between the metal and the semiconductor $R_c$ and the transport along the metal finger $R_s$ ~\cite{Steiner:2010id}. Given that the pixels can be asymmetric, these resistances need to be defined in both in-plane directions, $x$ and $y$:

\begin{align}
R_{sh}^x &= \frac{1}{A_r} R_{sh} \\
R_{sh}^y &= A_r R_{sh} \\
R_s^x &= \frac{1}{hA_r} \rho_m \\
R_s^y &= \frac{A_r}{h} \rho_m \\
R_c &= R_{back} = \frac{1}{L_x^2 A_r} \rho_c
\end{align}   

\noindent where $h$ is the height of the metal, $\rho_m$ their linear resistivity and $\rho_c$ the contact resistivity between metal and semiconductor. The sheet resistance of a stack of semiconductor layers $R_{sh}$ is equal to the combination in parallel of the individual sheet resistances. Using the single junction example of Fig. \ref{fig:spice_overview}, $R_{sh}(top)$ will be given by:

\begin{equation}
\frac{1}{R_{sh}(top)} = \frac{1}{R_{sh}(window)} + \frac{1}{R_{sh}(emitter)}
\end{equation}

Each of these can be estimated from the thickness of the layer $d$, the majority carrier mobility $\mu$ and the doping $N$ as ~\cite{Steiner:2011jy}:

\begin{equation}
\frac{1}{R_{sh}} = qd\mu N
\end{equation}

If the solar cell has been defined using only the DA and PDD junction models, this information is already available for all the layers of the structure. For junctions using the DB and two diode models, $R_{sh}$ will need to be provided for the top and bottom regions of each junction. Intrinsic layers will be ignored as they do not contribute to the lateral current transport.

\subsection{Solar array model}
\begin{figure}[h]
  \centering
    \includegraphics[width=0.9\columnwidth]{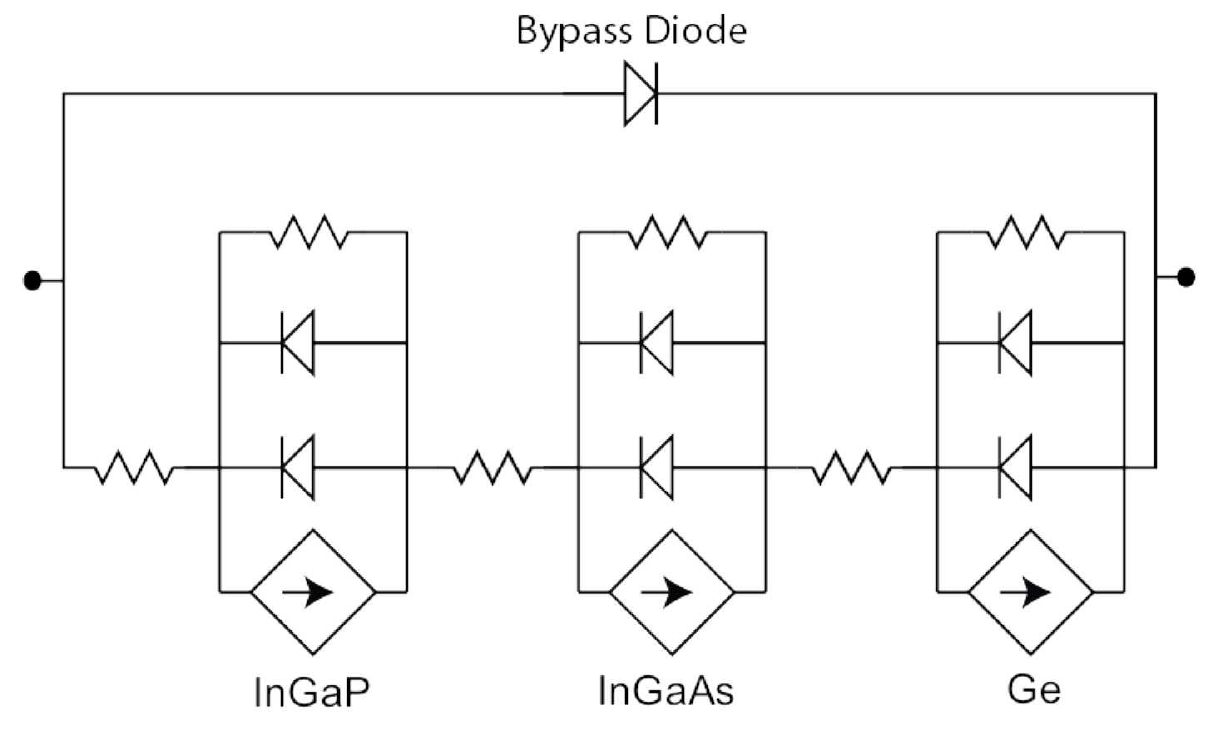}
  \caption{Equivalent circuit for a triple junction solar cell.}	
  \label{fig:3J_equiv_curcuit}
\end{figure}

The ability to use Solcore to build a SPICE equivalent circuit allows entire PV systems to be simulated from the bottom up~\cite{pspice}. Each photovoltaic solar cell is described using an equivalent circuit which can then be arranged in strings of series and parallel cells to represent the entire system. An example for a triple junction solar cell, complete with a bypass diode is shown in figure~\ref{fig:3J_equiv_curcuit}; this unit is the basic building block for a concentrator PV module~\cite{ekins:04b}.  

The diode and resistance values for the equivalent circuit are determined from solar cell testing, while the current source is evaluated by integrating the product of the spectral irradiance (estimated using an appropriate radiative transfer code e.g. SPCTRL2 or SMARTS) and the quantum efficiency which in turn can be calculated dynamically as a function of temperature by Solcore~\cite{ekins:ieee05}.  

Since the entire module (and subsequently the system) is assembled from individual solar cell components, it is possible (and indeed, necessary) to distribute the component values to accommodate for manufacturing tolerances. This enables a close match between the modelled output power and that measured experimentally and has been used to determine how both aerosols and precipitable water affect the electricity yield from concentrator PV systems~\cite{Chan:2012dm,Chan:2014dh}. Where system IV data is available, the emergence of electrical faults, (e.g. shunts or shading) can also be accounted for~\cite{kamath:2017}.  

\section{Closing remarks}

In this article we have described the capabilities of Solcore, a multi-scale, Python-based, modular simulation framework for semiconductor materials and solar cells. Its main strengths are:

\begin{itemize}
\item Flexibility: Provides a variety of tools, rather than a single solution, for the study of traditional and novel semiconductor materials and devices.
\item Modularity: Can be expanded with new capabilities, innovative solvers and tools.
\item Accessibility: Not only is it open source, but it is also designed to be easy to learn and to use, serving as a teaching tool as much as a research tool. 
\item Rigour: The physics behind every functionality are well understood and supported by numerous references, as are the approximations made in order to simplify the implementation of the problem or the interpretation of the results. 
\item Integrated: All of Solcore's features are designed to be compatible with one another to allow for truly multi-scale modelling in an integrated way.

\end{itemize}

\section*{Acknowledgements}

Solcore is not the creation of a single person but the combined effort of present and past members of the Quantum Photovoltaics Group at Imperial College London. Therefore, the authors would like to thank T. Thomas, N. L. A. Chan, J. Nelson and J. Connolly, who also contributed significantly to its development. The authors will also like to acknowledge the financial support of the Engineering and Physical Science Research council (EPSRC) through the research grant EP/M025012/1.

%\section*{References}

%\bibliographystyle{IEEEtran}  
%\bibliography{temp}


\end{document}